\documentclass[10pt,conference,letter]{IEEEtran}

\IEEEoverridecommandlockouts
\usepackage{color}
\usepackage{amsfonts}
\usepackage{colortbl}
\usepackage{framed}
\usepackage{balance}
\usepackage{verbatim}
\usepackage{fancyvrb}
\usepackage{listings}
\usepackage{graphicx}
\usepackage{bmpsize}
\usepackage{booktabs}
\usepackage{caption}
\usepackage{siunitx}
\usepackage{pifont}
\usepackage{amsmath}
\usepackage{multirow}
\usepackage{subfigure}
\usepackage{mathtools}
\usepackage{fancyhdr}
\usepackage{enumitem}
\usepackage{graphics}
\usepackage{fixltx2e}
\usepackage[algo2e]{algorithm2e}
\usepackage{physics}
\usepackage{mathtools}
\usepackage[hyphens]{url}

\usepackage{algorithm}
\usepackage[noend]{algpseudocode}

\usepackage{tikz}
\usetikzlibrary{arrows}
\usepackage[english]{babel}
\usepackage[utf8x]{inputenc}
\usepackage{amsmath}
\usetikzlibrary{arrows,automata}

\definecolor{dkgreen}{rgb}{0,0.6,0}
\definecolor{gray}{rgb}{0.5,0.5,0.5}
\definecolor{mauve}{rgb}{0.58,0,0.82}

\lstdefinestyle {c-lang}{frame=htb,
  language=C,
  aboveskip=3mm,
  belowskip=3mm,
  showstringspaces=false,
  columns=flexible,
  basicstyle={\footnotesize\ttfamily},
  numbers=left,
  numberstyle={\small},
  numbersep=5pt,
  xleftmargin=0.3cm,
  xrightmargin=0.3cm,
  rulesepcolor=\color{gray},
  numberstyle=\tiny\color{black},
  keywordstyle=\color{blue},
  commentstyle=\color{dkgreen},
  stringstyle=\color{mauve},
  breaklines=true,
  breakatwhitespace=true,
  tabsize=2,
  frame=shadowbox,
  captionpos=b
}
\lstdefinestyle {python-lang}{frame=htb,
  language=python,
  aboveskip=3mm,
  belowskip=3mm,
  showstringspaces=false,
  columns=flexible,
  basicstyle={\footnotesize\ttfamily},
  numbers=left,
  numberstyle={\small},
  numbersep=5pt,
  xleftmargin=0.3cm,
  xrightmargin=0.3cm,
  rulesepcolor=\color{gray},
  numberstyle=\tiny\color{black},
  keywordstyle=\color{blue},
  commentstyle=\color{dkgreen},
  stringstyle=\color{mauve},
  breaklines=true,
  breakatwhitespace=true,
  tabsize=2,
  frame=shadowbox,
  captionpos=b
}

\newcommand{\mpifordask}[0]{\textsf{\small MPI4Dask }}
\newcommand{\mpiforpy}[0]{\textsf{\small mpi4py }}
\newcommand{\mvgdr}[0]{\textsf{\small MVAPICH2-GDR }}
\newcommand{\daskmpi}[0]{\textsf{\small Dask-MPI }}
\newcommand{\mpifordaskns}[0]{\textsf{\small MPI4Dask}}
\newcommand{\mpiforpyns}[0]{\textsf{\small mpi4py}}
\newcommand{\mvgdrns}[0]{\textsf{\small MVAPICH2-GDR}}
\newcommand{\daskmpins}[0]{\textsf{\small Dask-MPI}}

\newcommand{\indentcentercol}[0]{\indent\indent\indent\indent\indent\indent\indent\indent\indent}

\newcommand{\codesf}[1]{\textsf{\small #1}}

\usepackage[
square,
numbers,
comma,
sort&compress
]
{natbib}

\makeatletter
\newcommand\etc{etc\@ifnextchar.{}{.\@}}
\makeatother

\newcolumntype{K}{\columncolor[gray]{0.8}\raggedright}

\newcommand{\MySection}[1]{
  \section{#1}
}
\newcommand{\MySubsection}[1]{
  \vspace{-0.5ex}
  \subsection{#1}
  \vspace{-0.5ex}
}
             
\newcommand{\MySubsubsection}[1]{
  \subsubsection{#1}
}

\newcommand{\MyCaption}[1]{
  \caption{#1}
}

\begin{document}
\title{Efficient MPI-based Communication for GPU-Accelerated Dask Applications\\
\thanks{This research is supported in part by NSF grants \#1818253, \#1854828, \#1931537, \#2007991, \#2018627, and XRAC grant \#NCR-130002.}
}

\author{
	Aamir Shafi, Jahanzeb Maqbool Hashmi, Hari Subramoni and Dhabaleswar K. (DK) Panda \\
	The Ohio State University\\
    
    \begin{normalsize}
        \begin{sffamily}
            \{shafi.16, hashmi.29, subramoni.1, panda.2\}@osu.edu
        \end{sffamily}
    \end{normalsize}
 
}

\maketitle
\pagestyle{plain}


\begin{abstract}


Dask is a popular parallel and distributed computing framework, which rivals Apache Spark to enable task-based scalable processing of big data. The Dask Distributed library forms the basis of this computing engine and provides support for adding new communication devices. It currently has two communication devices: one for TCP and the other for high-speed networks using UCX-Py---a Cython wrapper to UCX. This paper presents the design and implementation of a new communication backend for Dask---called MPI4Dask---that is targeted for modern HPC clusters built with GPUs. MPI4Dask exploits mpi4py over MVAPICH2-GDR, which is a GPU-aware implementation of the Message Passing Interface (MPI) standard. MPI4Dask provides point-to-point asynchronous I/O communication {\em coroutines}, which are non-blocking concurrent operations defined using the {\tt async}/{\tt await} keywords from the Python's {\tt asyncio} framework. Our latency and throughput comparisons suggest that MPI4Dask outperforms UCX by $6\times$ for 1 Byte message and $4\times$ for large messages (2 MBytes and beyond) respectively. We also conduct comparative performance evaluation of MPI4Dask with UCX using two benchmark applications: 1) sum of cuPy array with its transpose, and 2) cuDF merge. MPI4Dask speeds up the overall execution time of the two applications by an average of $3.47\times$ and $3.11\times$ respectively on an in-house cluster built with NVIDIA Tesla V100 GPUs for $1-6$ Dask workers. We also perform scalability analysis of MPI4Dask against UCX for these applications on TACC's Frontera (GPU) system with upto $32$ Dask workers on $32$ NVIDIA Quadro RTX 5000 GPUs and $256$ CPU cores. MPI4Dask speeds up the execution time for cuPy and cuDF applications by an average of $1.71\times$ and $2.91\times$ respectively for $1-32$ Dask workers on the Frontera (GPU) system. 
\end{abstract}


\MySection{Introduction}
\label{sec:intro}

With the end of Moore's Law~\cite{moore:1965, Moore75Reprint} in sight, the performance advances in the computer industry are likely to be driven by the ``Top''---1) hardware architecture, 2) software, and 3) algorithms---as noted by Leiserson recently~\cite{Leisersoneaam9744}. This is in stark comparison to the vision ``There's Plenty of Room at the Bottom'' laid out by Feynman~\cite{Feynman} in 1959 referring to semiconductor physics and fabrication technology. Leiserson~\cite{Leisersoneaam9744} argues that the post-Moore generation of software will focus on reducing software engineering bloat as well as exploiting parallelism, locality, and heterogeneous hardware. A representative of the ``Top'' is the Python programming language, which is a clear winner on the landscape of data science. Python is a classic example of a language benefiting from Moore's law by traditionally focusing on programmer productivity and reduced development time. In the context of data science applications, Python's popularity is due to rich set of free and open-source libraries that enable the end-to-end data processing pipeline. These libraries/packages include: core (SciPy), data preparation (NumPy, Pandas), data visualization (Matlibplot), machine learning (Scikit-learn), and deep learning (PyTorch, TensorFlow). However in the post-Moore era, it is vital that Python is able to support parallel and distributed computing as well as exploit emerging architectures especially accelerators.

Two popular big data computing frameworks that enable high-performance data science in Python include Dask~\cite{matthew_rocklin-proc-scipy-2015} and Apache Spark~\cite{ZahariaXinEtAl16cacm}. This paper focuses on Dask, which provides support for natively extending popular data processing libraries like numPy and Pandas. This allows incremental development of user applications and lesser modifications to legacy codes when executed with Dask. Traditionally Dask has mainly supported execution on hosts (CPUs) only, which means that it was not able to exploit massive parallelism offered by Graphical Processing Units (GPUs). However this has recently changed as part of the development of the NVIDIA RAPIDS library, which aims to enable parallel and distributed computation on clusters of GPUs. RAPIDS adopted a similar approach---as taken by Dask---of extending already existing Python data processing libraries for the GPU ecosystem. For instance, RAPIDS support processing of distributed data stored in cuPy (numPy-like) and cuDF (Pandas-like) formats. An overarching goal for RAPIDS project is to hide the low-level programming complexities of the CUDA compute environment from Python developers and make it easy to deploy and execute full end-to-end processing pipeline on GPUs. An example of a machine learning library provided by RAPIDS is cuML~\cite{Raschka2020MachineLI}, which is the GPU-counterpart for Scikit-learn. 

In order to support execution of Dask programs on cluster of GPUs, an efficient communication layer is required. This is extended by the Dask Distributed library that provides essentials for distributed execution of Dask programs on parallel hardware. It is an asynchronous I/O application, which means that it supports non-blocking and {\em concurrent} execution of its routines/functions including communication primitives. This mandates the restriction that any communication backend---aiming to be part of Dask Distributed library---must implement {\em coroutines} that are non-blocking methods defined and awaited by using the {\tt async} and {\tt await} keywords respectively. Coroutines support concurrent mode of execution and hence cannot be invoked like regular Python functions. The main reason behind this non-blocking asynchronous style of programming is to avoid blocking, or delaying, networking applications for completing I/O operations. 

The Dask Distributed library currently provides two communication devices: one for TCP and the other for UCX-Py that is a Cython-based wrapper library to UCX~\cite{UCX}. The UCX-Py communication device is capable of communicating data to/from GPU device memory directly. However, it fails to deliver high-performance---as revealed by our performance evaluation detailed in Section~\ref{sec:results}---since it delays progressing the communication engine by assigning it to a separate coroutine that executes periodically. We address this performance overhead by advocating an efficient design where communication coroutines also progress the communication engine.

The Message Passing Interface (MPI) standard~\cite{MPI123} is considered as the {\em defacto} programming model for writing parallel applications on modern GPU clusters. The \mvgdr~\cite{MVAPICH2-GDR} library provides high-performance support for communicating data to/from GPU devices via optimized point-to-point and collective routines. As part of this paper, we design, implement, and evaluate a new communication backend, called \mpifordaskns, based on the \mvgdr library for Dask. \mpifordask implements point-to-point communication coroutines using \mpiforpy~\cite{DALCIN20111124} over \mvgdrns. To the best of our knowledge, this is the first attempt to use an GPU-enabled MPI library to handle communication requirements of Dask. This is challenging especially because the Dask Distributed library is an asynchronous I/O application. Hence any point-to-point communication operations---implemented via MPI---must be integrated as communication coroutines that typically have performance penalties over regular Python functions. An additional goal here is to avoid making unnecessary modifications to the Dask ecosystem.

Another challenge that \mpifordask addresses is to provide communication isolation and support for dynamic connectivity between Dask components including scheduler, workers, and client. Communication between Dask entities---as provided by TCP and UCX backends---is based on the abstraction of endpoints that are established when processes connect with one another. An endpoint essentially represent a direct point-to-point connection. In \mpifordask, we devise a strategy, detailed in Section~\ref{sec:impl}, that relies on sub-communicator mechanism provided by MPI---coupled with communicator duplication---to handle this. This allows us to build sub-communicators between processes to mimic a direct endpoint-based connection. 

\begin{figure*}[htbp]
    \centering
    \vspace{-1.5ex}
    \subfigure[\codesf{Latency (Small)}]
    {
        \label{fig:ri2-pingpong-lat-small-intro}
        \includegraphics[width=0.32\textwidth]{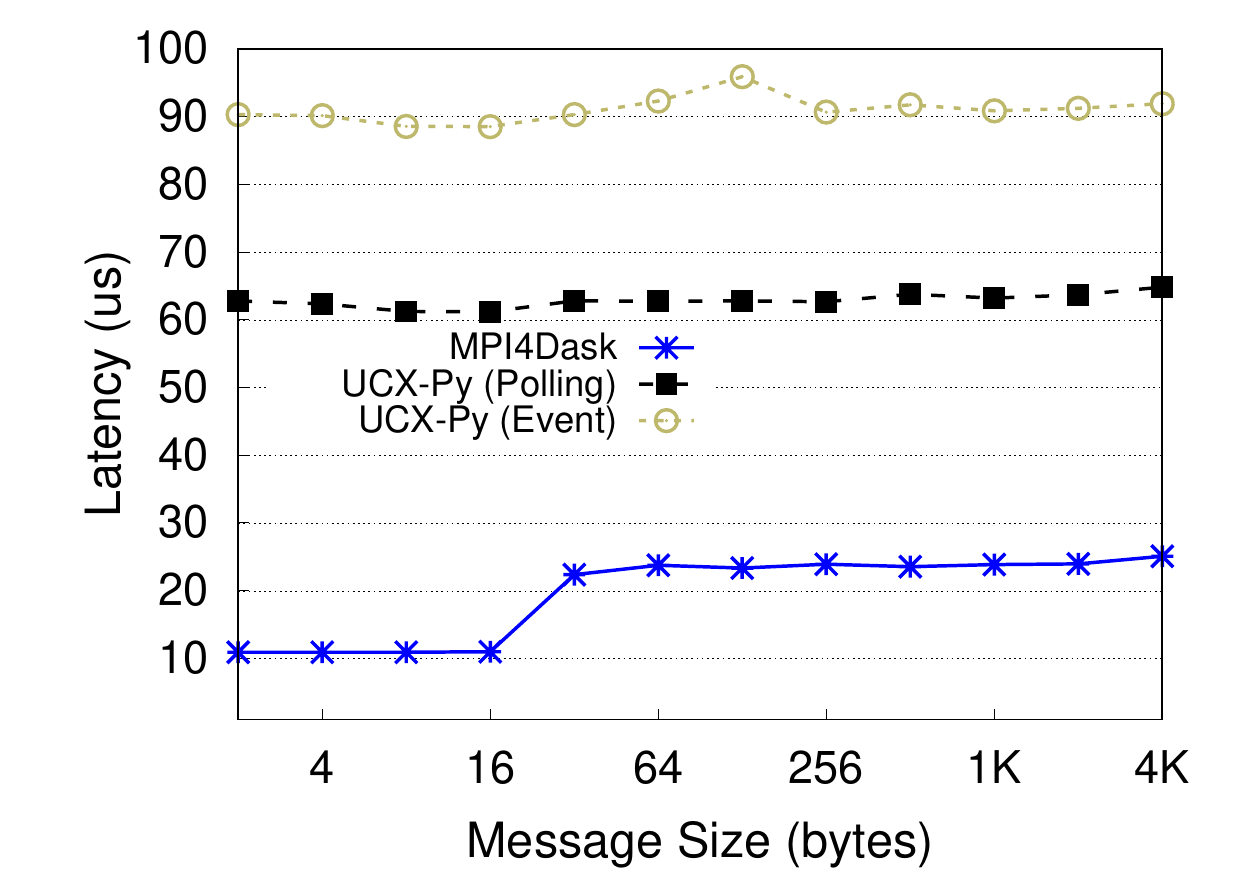}
    }
    \hspace{-3ex}
    \subfigure[\codesf{Latency (Medium)}]
    {
        \label{fig:ri2-pingpong-lat-med-intro}
        \includegraphics[width=0.32\textwidth]{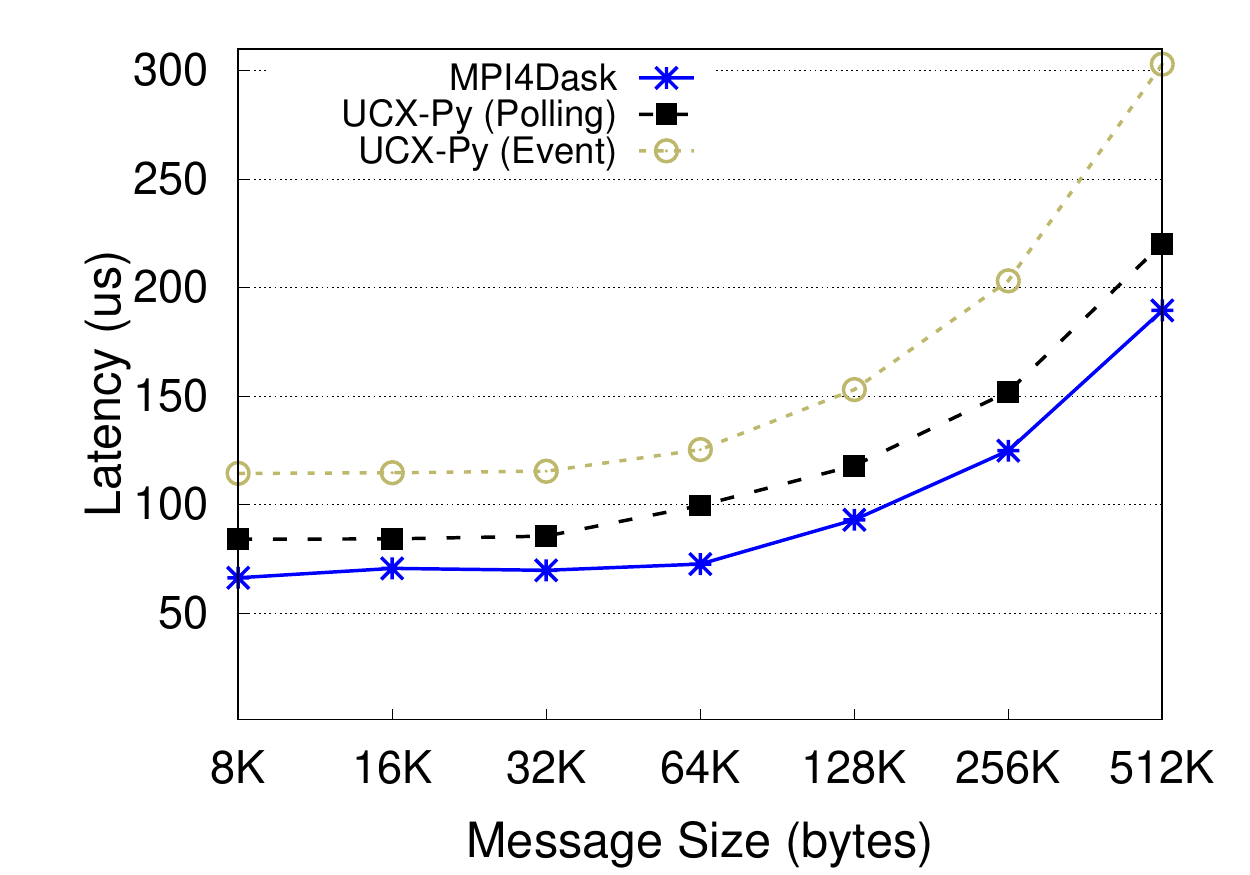}
    }
    \hspace{-3ex}
    \subfigure[\codesf{Throughput (Large)}]
    {
        \label{fig:ri2-pingpong-bw-large-intro}
        \includegraphics[width=0.32\textwidth]{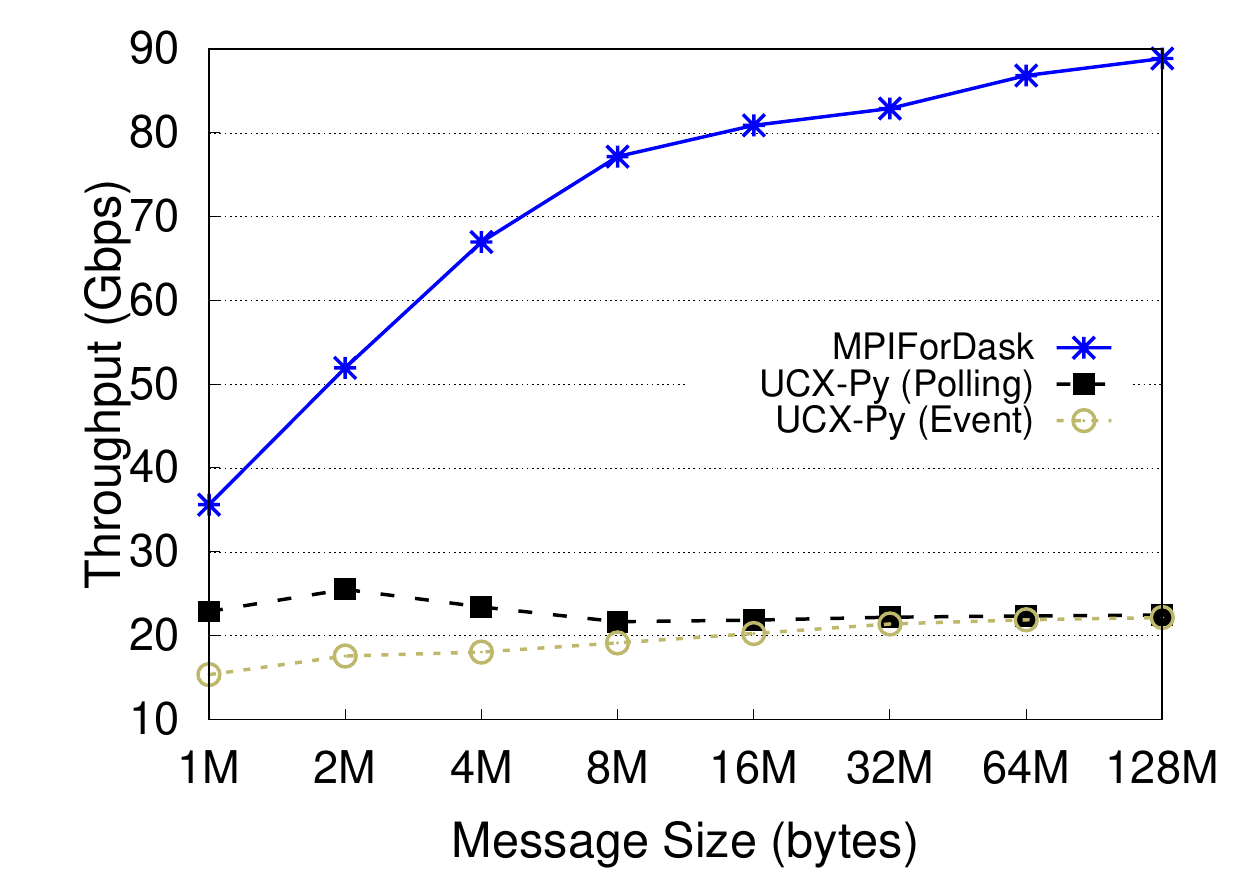}
    }
    \vspace{-1ex}
    \MyCaption{Latency and Bandwidth comparison of \mvgdr (using \mpiforpyns) with UCX using Ping-pong Benchmark---based on Python Coroutines---on the RI2 Cluster (V100 GPUs). This demonstrates the performance benefits of of using \mvgdr as compared to UCX at the Python layer and forms the major motivation of this paper.}
    \vspace{-2ex}
    \label{fig:ri2-pingpong-latbw-introduction}
\end{figure*}

In order to motivate the need for \mpifordaskns, we present latency comparison between Python communication coroutines implemented using \mvgdr and \mpiforpy with UCX-Py polling and event-based modes in Figure~\ref{fig:ri2-pingpong-latbw-introduction}. The latency graphs---Figure~\ref{fig:ri2-pingpong-lat-small-intro} and Figure~\ref{fig:ri2-pingpong-lat-med-intro} depict that the performance of Python communication coroutines using \mvgdr (with \mpiforpyns) are roughly $5\times$ better for small messages ($1$ Byte to $16$ Bytes) and $2-3\times$ better for small-medium messages ($32$ Bytes to $4$ KBytes). For large messages ($2$ MBytes to $128$ MBytes) as shown in Figure~\ref{fig:ri2-pingpong-bw-large-intro}, \mvgdr (with \mpiforpyns)-based coroutines are better than UCX-Py point-to-point coroutines by a factor of $3-4\times$. These performance benefits form the basis for our motivation to design and implement MPI-based communication backend for the Dask framework.

We present a detailed performance evaluation of \mpifordask against TCP and UCX communication backends using a number of micro-benchmarks and application benchmarks---these are presented in Section~\ref{sec:results}. For the micro-benchmark evaluation, \mpifordask outperforms the UCX communication device in latency and throughput comparison by ping-pong benchmarks by $5\times$ for small messages, $2-3\times$ better for small-medium messages, and  $3-4\times$ for large messages. 
We used these two application benchmarks: 1) sum of cuPy array and its transpose, and 2) cuDF merge on an in-house cluster and the TACC's Frontera (GPU) cluster. On the in-house cluster, we are witnessing an average speedup of $3.47\times$ for $1-6$ Dask workers for the cuPy application. Communication time for this application has been reduced by $6.92\times$ compared to UCX communication device. For the cuDF application on the in-house cluster, there is an average speedup of $3.11\times$ for $2-6$ Dask workers and $3.22\times$ reduction in communication time. On the Frontera (GPU) cluster, \mpifordask outperforms UCX by an average factor of $1.71\times$ for the cuPy application with $1-32$ Dask workers. For the cuDF application, \mpifordask reduces the total execution time by an average factor of $2.91\times$ when compared to UCX for $1-32$ Dask workers. Reasons for better overall performance of \mpifordask against its counterparts include better point-to-point performance of \mvgdr and efficient coroutine implementation. Unlike UCX-Py that implements a separate coroutine to make progress for UCX worker, \mpifordask ensures {\em cooperative} progression where every communication coroutine triggers the communication progression engine.

\MySubsection{Contributions}
\label{sec:intro:contribution}

Main contributions of this paper are summarized below: 

\begin{enumerate}

    \item Design and implementation of \mpifordask that provides high-performance point-to-point communication coroutines for Python-based HPC applications. To the best of our knowledge, \mpifordask is the first library that implements MPI-based Python common coroutines that work with the {\tt asyncio} framework on cluster of GPUs.
    
    \item Integration of \mpifordask with {\em asynchronous} Dask Distributed library. This is a pioneering effort that enables MPI-based communication for the Dask ecosystem.
    
    \item Demonstrate the performance benefits of using \mpifordask compared to TCP and UCX devices using basic micro-benchmarks and two application benchmarks (based on cuPy and cuDF) using an in-house cluster comprising of V100/K80 GPUs.
    
    \item Perform scalability evaluation of \mpifordask against TCP and UCX communication devices using two application benchmarks (based on cuPy and cuDF) on TACC's Frontera (GPU) system with upto $32$ Dask workers on $32$ NVIDIA Quadro RTX 5000 GPUs and $256$ CPU cores.
    
\end{enumerate}

Rest of the paper is organized as follows. Section~\ref{sec:back} presented relevant background followed by the design approach of the \mpifordask library in Section~\ref{sec:design}. Section~\ref{sec:impl} presents implementation details. Later, we evaluate \mpifordask using a communication micro-benchmark and two application-level benchmarks. Section~\ref{sec:conclusion} concludes the paper.
\MySection{Background}
\label{sec:back}

This section covers the background for this paper, which includes the fundamentals of the MPI standard, the Dask framework, and the {\tt asyncio} package. This section also presents related work.

\MySubsection{Message Passing Interface (MPI)}

The Message Passing Interface (MPI) API is considered the {\em defacto} standard for writing parallel applications. The MPI API defines a set of point-to-point and collective communication routines that are provided as convenience functions to application developers. 
In the context of the Dask Distributed library, the most relevant set of functions are the non-blocking point-to-point communication functions like {\tt MPI\_Isend()} and {\tt MPI\_Irecv()}. Both of these functions return an {\tt MPI\_Request} object that can be used to invoke {\tt MPI\_Test()} function to check if the non-blocking communication operation has completed or not. In this paper, we make use of a GPU-aware MPI library called \mvgdr \cite{MVAPICH2-GDR} that provides optimized point-to-point and collective communication support for GPU devices. GPU-awareness here means that the MPI library is capable of directly communication data efficiently to/from GPU memory instead of staging it through the host system. The Dask ecosystem is implemented in the Python programming language. This implies that the MPI communication backend must also be implemented in Python. For this reason, we use the GPU-aware \mpiforpy library that provides Cython \cite{behnel2011cython} wrappers to native MPI library---\mvgdr in this case. For performance reasons, it is important that the Python wrapper library is capable of communicating data to/from GPU memory directly without incurring the overhead of serialization/de-serialization. \mpiforpy supports efficient exchange of GPU data stored in cuPy and cuDF format.

\MySubsection{Dask}

%

\begin{figure}[htbp]
    \centering
    \includegraphics[width=0.8\linewidth]{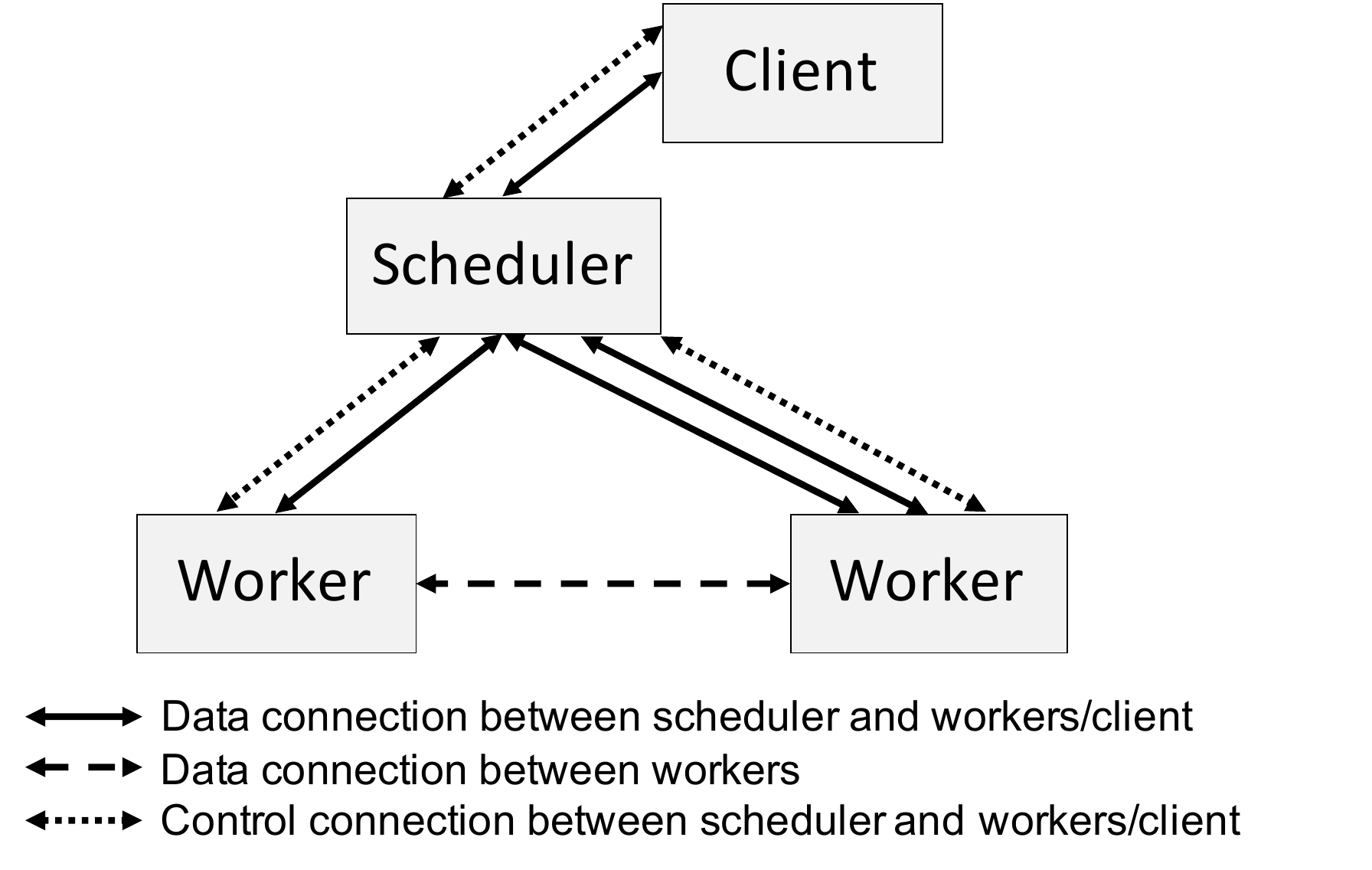}
    \caption{Dask Execution Model. The scheduler and workers form a Dask cluster. The client executes user program by connecting with the Dask cluster.
    }
    \label{fig:impl:dask-comms}
    \vspace{-1ex}
\end{figure}

Dask is a popular data science framework for Python programmers. It converts user application into a task-graph, which is later executed {\em lazily} on distributed hardware. This execution requires data exchange supported through implicit communication by the Dask runtime. The Dask ecosystem is a suite of Python packages. One such package is Dask Distributed that provides various components like scheduler, workers, and client. This library also supports point-to-point functionality between these Dask components. Figure \ref{fig:impl:dask-comms} depicts the distributed execution model of a Dask program. There are three types of connections between Dask entities: 1) data connections (solid lines) for exchanging application data, 2) control connections (dotted lines) for exchanging heart-beat messages to maintain status of workers and detect any failures, and 3) dynamic connections (dashed lines) between workers to resolve dependencies, work-stealing, or achieving higher throughput. Currently Dask has two communication devices: 1) the TCP device that exploits the asynchronous Tornado library, 2) the UCX backend that uses UCX-Py \cite{ucx-py}---a Cython wrapper library---on top of the native UCX \cite{UCX} communication library. 

\subsection{The {\tt asyncio} Package}

Since version $3.5$, Python has introduced a new package called {\tt asyncio} that allows writing concurrent non-blocking I/O applications. The main idea behind this package is to allow networking (or other I/O) applications to efficiently utilize CPU without unnecessarily getting blocked for long-running I/O operations. This is possible because as the Python program executes, it keeps defining {\em tasks} that are stored in a task queue and execute concurrently as soon as it is possible to execute them. These tasks are defined through coroutines, which are functions defined using the {\tt async} keyword and invoked later using the {\tt await} keyword. 

\MySection{Design Overview of \mpifordask}
\label{sec:design}

This section presents the design of the \mpifordask library. We first discuss communication requirements of the Dask framework. This is followed by coverage of layered architecture of Dask with focus on communication devices.

We begin with communication requirements of the Dask framework.

\begin{enumerate}
    \item {\textbf{Scalability:}} Provide scalability by exploiting low-latency and high-throughput for cluster of GPUs typically deployed in modern data centers. 
    \item {\textbf{Coroutines:}} The communication backend for Dask Distributed library is asynchronous and executes within an event loop, as part of an {\tt asyncio} application. Hence, the communication backend needs to support non-blocking point-to-point send and receive operations through {\tt asyncio} coroutines defined using the {\tt async}/{\tt await} syntax.
	\item {\textbf{Elasticity:}} The Dask cluster (e.g., scheduler and a set of workers) is an elastic entity meaning that the workers dynamically join or leave the cluster. In this context, the communication backend needs to support such dynamic connectivity for clients and workers. Dask, however, also executes with static number of processes using the \daskmpi library. In this paper we adopt this approach.
	\item {\textbf{Serialization/De-serialization:}} Dask applications support distributed processing of GPU-based buffers including cuPy and cuDF DataFrames. The communication engine should be able to support communication to/from GPU-based Python objects/data-structures supported by Dask.
\end{enumerate}

\begin{figure}[htbp]
\vspace{-1ex}
    \centering
    \includegraphics[width=0.95\linewidth]{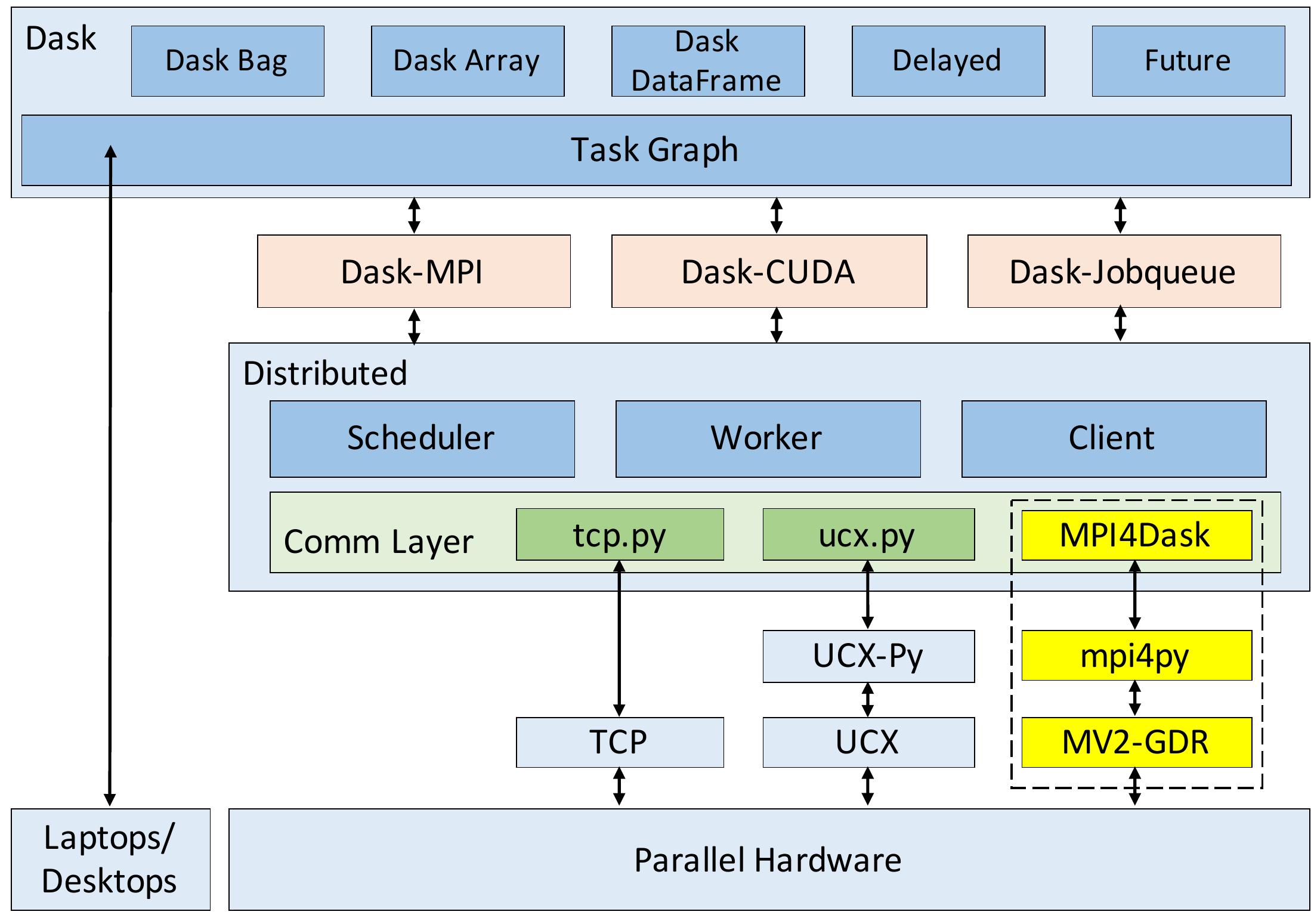}
    \caption{Dask Layered Architecture with Communication Backends. Yellow boxes are designed and evaluated as a part of this paper.
    }
    \label{fig:dask-layered-arch}
\vspace{-1ex}
\end{figure}

Figure \ref{fig:dask-layered-arch} presents a layered architecture for Dask. The top layer represents the Dask framework, which includes support for various data structures and storage formats including Dask Bags, Arrays, and DataFrames. Also there is support for asynchronous execution through {\tt Delayed} and {\tt Future} objects. The user application is internally converted to a task graph by Dask. The next layer shows various packages that are part of the Dask ecosystem. These include \daskmpins, {\tt Dask-CUDA}, and {\tt Dask-Jobqueue}. The next layer represents the Dask Distributed library. This package supports distributed computation through scheduler, worker, and client. This library also contains the {\tt distributed.comm} module, shown here as ``Comm Layer''. This layer provides the API that all Dask communication backends must implement. A subset of this API can be seen in Listing \ref{list:backendapi}. As shown, the Dask Distributed library provides TCP and UCX-Py backends represented in this layer by {\tt tcp.py} and {\tt ucx.py}. \mpifordask is implemented as part of the Dask Distributed ``Comm Layer'' as a multi-layered software sitting over \mpiforpyns, which in turn exploits \mvgdr for GPU-aware MPI communication. The bottom layer in Figure \ref{fig:dask-layered-arch} represents cluster of GPUs as parallel hardware. Dask is also capable of running on shared memory system like a laptop or a desktop.

\begin{figure}[htbp]
\centering
\begin{lstlisting}[style=python-lang, label=list:backendapi,caption={Subset of the Communication Backend API as Mandated by the Dask Distributed Library}, captionpos=b]
class Comm(ABC):
    @abstractmethod
    def read(self, deserializers=None):
    @abstractmethod
    def write(self, msg, serializers=None, 
                           on_error=None):

class Listener(ABC):
    @abstractmethod
    async def start(self):
    @abstractmethod
    def stop(self):

class Connector(ABC):
    @abstractmethod
    def connect(self, address, deserialize=True)
    
\end{lstlisting}
\vspace{-4ex}
\end{figure}
\MySection{Implementation Details of the \mpifordask Library}
\label{sec:impl}

This section outlines the implementation details of the \mpifordask communication library. 

\subsection{Bootstrapping the Dask ecosystem and initializing MPI}
\label{sec:impl:bootstrapping}

There are many ways to start execution of Dask programs. The manual way is to start the scheduler followed by multiple workers on the command line. After this, the client is ready to connect with the Dask cluster and execute the user program. In order to automate this process, Dask provides a number of utility ``Cluster'' classes. An example of this is the {\tt LocalCUDACluster} from the Dask CUDA package. Once an instance of {\tt LocalCUDACluster} object has been initiated, a client can connect to execute a user program. Other example of utility cluster classes include {\tt SLURMCluster}, {\tt SGECluster}, {\tt PBSCluster}, and others.

\mpifordask currently requires that user initiates the execution of Dask program using the \daskmpi package. \daskmpi uses the bootstrapping mechanism provided by MPI libraries to start Dask scheduler, client, and one or more workers. We use the {\tt mpirun\_rsh} utility provided by \mvgdr. Figure~\ref{fig:impl:dask-comms} illustrates this where the {\tt mpirun\_rsh} utility was used to start $4$ MPI processes. Here processes $0$ and $1$ assume the role of scheduler and client, while all the other processes---in this case $2-3$ become worker processes. This bootstrapping is done by the \daskmpi package and this is the point where \mpiforpy and \mvgdr are also initialized. The {\tt CUDA\_VISIBLE\_DEVICES} environment variable is used to map Dask worker processes on a particular GPU in a node. This is particularly important when a node has multiple GPUs since we would want multiple Dask workers to utilize distinct GPUs on the system. Note that \daskmpi does not provide any communication between Dask components and this is what we address in this paper.



\subsection{Point-to-point Communication Coroutines}
\label{sec:impl:pt2pt}

A challenge that we tackle as part of this work is to implement asynchronous communication coroutines using \mpiforpy over \mvgdr that can be incorporated inside the Dask Distributed layer. To the best of our knowledge, this is first such effort to exploit MPI-based communication inside an {\tt asyncio} application in Python.

\mpiforpy provides two variants of point-to-point functions. The first option is to use the {\em lowercase} methods like {\tt Comm.send()}/{\tt Comm.isend()} to communicate data/to from Python objects. This involves picking/unpickling (serialization/de-serialization) of Python objects. The second option is to use methods like {\tt Comm.Send()}/{\tt Comm.Isend()} that start with {\em uppercase} letter and communicate data to/from directly from user-specified buffer. The second option is efficient and preferred for achieving high-performance in the Dask Distributed library. Since we are implementing \mpifordask for GPU buffers, the specified buffer to \mpiforpy communication routines must support the {\tt \_\_cuda\_array\_interface\_\_}.

As previously mentioned, the Dask Distributed library is an asynchronous application due to which it executes coroutines in a non-blocking and concurrent manner. Also for this reason, using blocking MPI point-to-point methods will result in deadlock for the application. In order to tackle this, our implementation of \mpifordask makes use of {\tt Comm.Isend()}/{\tt Comm.Irecv()} methods that return {\tt Request} objects. Later, \mpifordask checks for completion of pending communication using the {\tt Request.Test()} method. Instead of checking the completion status in a busy-wait loop---that will result in a deadlock situation---\mpifordask calls the {\tt asyncio.sleep()} method that allows other coroutines to make progress while waiting for communication to complete. Listings \ref{list:sendcoro} and \ref{list:recvcoro} provide outlines of send and receive communication coroutines in \mpifordaskns. 

\begin{figure}[htbp]
\centering
\begin{lstlisting}[style=python-lang, label=list:sendcoro, caption={The {\tt Comm.Isend()}-based Send Communication Coroutine Implemented by \mpifordaskns.}, captionpos=b]
request = comm.Isend([buf, size], dest, tag)
status = request.Test()

while status is False:
    await asyncio.sleep(0)
    status = request.Test()
\end{lstlisting}
\vspace{-6ex}
\end{figure}

\begin{figure}[htbp]
\centering
\begin{lstlisting}[style=python-lang, label=list:recvcoro, caption={The {\tt Comm.Irecv()}-based Receive Communication Coroutine Implemented by \mpifordaskns.}, captionpos=b]
request = comm.Irecv([buf, size], src, tag)
status = request.Test()

while status is False:
    await asyncio.sleep(0)
    status = request.Test()
\end{lstlisting}
\vspace{-6ex}
\end{figure}

\subsection{Handling Large Messages}
\label{sec:impl:large}

Dask is a programming framework for writing data science applications and it is common for such libraries to handle and exchange large amounts of data. As a consequence it is possible that the higher layers of the Dask ecosystem attempt to communicate large messages using the underlying communication infrastructure. We have experienced this with \mpifordask that relies on the point-to-point non-blocking primitives provided by MPI---{\tt Comm.Isend()} and {\tt Comm.Irecv()} methods. These functions accept an argument {\tt int count} that is used to specify the size of the message being sent or received. The maximum value that this parameter can hold is $2^{31}-1$ bytes, which corresponds a message size of $2$ GB$-1$. On the other hand, the Dask Distributed library attempts to communicate messages larger than this value including upto $64$ GB. We have catered for this requirement by dividing the large message into several chunks of $1$ GB for the actual communication using the {\tt Comm.Isend()} and {\tt Comm.Irecv()} methods. Typically the buffers specified to these communication functions are Python objects and hence subscriptable using the slice notation {\tt array[start:end]}. This approach works for numPy and cuPy arrays and hence the buffer argument can be subscripted in a loop to implement chunking. But this strategy does not work for cuDF and RAPIDS Memory Manager (RMM) buffers {\tt rmm.DeviceBuffer}. For these, we implement chunking by incrementing the offset argument in the communication loop.

\subsection{Providing Communication Isolation}
\label{sec:impl:isolation}

\begin{figure}[htbp]
    \centering
    \includegraphics[width=0.7\linewidth]{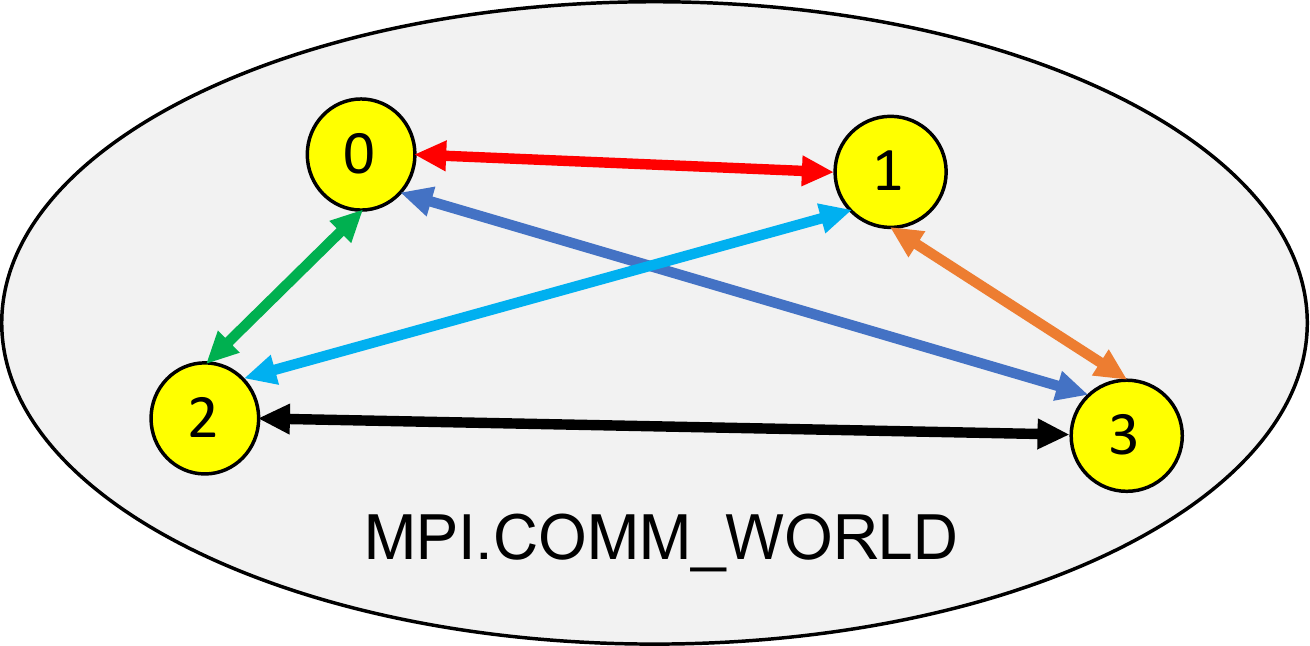}
    \caption{Building new Sub-Communicators from {\tt MPI.COMM\_WORLD} at Startup. The bi-directional edges between processes represent newly built sub-communicators.
    }
    \label{fig:impl:subcomms}
    \vspace{-1ex}
\end{figure}

After the initial bootstrapping, \mpifordask has full connectivity between all processes. This is provided by the default communicator {\tt MPI\_COMM\_WORLD} initialized by the MPI library. However using this communicator naively might lead to message interference. This is possible when a particular worker might want to use same tag for both data and control message exchanges. There are multiple ways to tackle this. The approach that we choose in this paper is to rely on MPI sub-communicators. Figure~\ref{fig:impl:subcomms} outlines this approach. The $4$ MPI processes here engage with one another to form five sub-communicators represented by colored bi-directional edges between a pair of processes. The pseudocode for this is shown in Listing~\ref{list:initloop}. Each process stores a handle to these sub-communicators in a table called {\tt comm\_table}. Creating a new MPI sub-communicator is a costly operation and for this reason all of this is done at the startup. The {\tt MPI.Group.Incl()} and {\tt MPI.COMM\_WORLD.Create()} functions (from \mpiforpyns) are used for creating sub-communicators that only contain two processes. Complexity of making new sub-communicators is $O(\frac{n^2}{2})$ where $n$ is the total number of MPI processes. Our approach has negligible overhead because these sub-communicators are initialized at the startup and re-used later during the program execution.

\begin{figure}[htbp]
\centering
\begin{lstlisting}[style=python-lang, label=list:initloop, caption={Nested Initialization Loop Executed at Each MPI Process during Startup to Build new Sub-communicators from {\tt MPI.COMM\_WORLD}. The newly built sub-communicators encapsulate only two processes and are stored in the {\tt comm\_table}.}, captionpos=b]
for i in range(size):
   for j in range(i+1, size):
       incls = [i, j]
       new_group = MPI.Group.Incl(group, incls)
       new_comm = MPI.COMM_WORLD.Create(new_group)
       if rank == i:
           comm_table.update({j : new_comm})
       else if rank == j:
           comm_table.update({i : new_comm})
\end{lstlisting}
\vspace{-6ex}
\end{figure}

\subsection{Handling Dynamic Connectivity}
\label{sec:impl:dynamic}

The Dask Distributed library support connectivity between Dask components including scheduler, workers, and client. The UCX and TCP communication devices maintain information for remote endpoints since this information is required for the actual communication. In \mpifordaskns, we have replaced the abstraction of endpoint with sub-communicator. This enables us to utilize the existing communication infrastructure of the Dask ecosystem. Apart from communications being setup at the startup, Dask allows dynamic connections between workers as well. In this context, new communication channels are only established in the Dask Distributed library through traditional client/server semantics of connect/accept. In \mpifordask we handle this requirement by starting a server that listens for incoming connections---and invokes a connection handler callback function---using the {\tt asyncio.start\_server()} method. Dask scheduler and worker processes use this function since they act as listeners and other processes are allowed to connect with them. Later any process (scheduler, client, or worker) that attempts connecting to a listener process do so by using the {\tt asyncio.open\_connection()} function. When two processes connect with one another---and this is something that happens frequently in Dask at startup---each process gets the relevant sub-communicator from the {\tt comm\_table} by doing a lookup based on the destination process rank in the {\tt MPI.COMM\_WORLD} communicator. The corresponding sub-communicator from the {\tt comm\_table} is {\em duplicated} by using the {\tt Comm.Dup()} function. This ensures that a new sub-communicator is built for every new dynamic connection. For performance reasons, we are maintaining a configurable cache of these sub-communicators.

\MySection{Performance Evaluation}
\label{sec:results}

This section presents performance evaluation of \mpifordask against UCX and TCP (using IPoIB) communication devices using 1) Ping Pong micro-benchmark (Section~\ref{sec:results:microbenchmark}), 2) Two application benchmarks (cuPy and cuDF) on an in-house cluster (RI2) with two types of GPU nodes with NVIDIA Tesla V100s and NVIDIA Tesla K80s (Section~\ref{sec:results:appbenchmark}), and 3) Scalability results for the same two application benchmarks on TACC's Frontera (GPU) cluster with NVIDIA Quadro RTX 5000 GPUs (Section~\ref{sec:results:scalability}). The hardware specifications for the RI2 cluster and Frontera (GPU) subsystem are shown in Table~\ref{tab:specs}. The following versions of the software were used: UCX v1.8.0, UCX-Py v0.17.0, Dask Distributed v2.30, MVAPICH2-GDR v2.3.4, and mpi4py v3.0.3.

\begin{table}[htbp]
\centering
    \renewcommand{\tabcolsep}{3pt}
\caption{Hardware specification of the in-house RI2 and TACC's Frontera (GPU) clusters. Columns 1 and 2 titled {\textsf{\small RI2-V100}} and {\textsf{\small RI2-K80}} provide details for two types of nodes on the RI2 cluster. Column 3 titled {\textsf{\small RI2-V100}} provide details for the nodes used on the Frontera (GPU) cluster.}
\label{tab:specs}
\begin{tabular}{@{} *5l}    \toprule
{\bf Specification}& {\bf RI2-V100}     &  {\bf RI2-K80}    & {\bf Frontra (GPU)} \\
\midrule
Number of Nodes     &   16              &   16              & 90\\
Processor Family	&	Xeon Broadwell	&	Xeon Broadwell	& Xeon Broadwell\\
Processor Model	    &	E5-2680 v4	    &	E5-2680 v4	    & E5-2620 v4 \\
Clock Speed	        &	2.4 GHz     	&	2.4 GHz	        & 2.1 GHz\\
Sockets	            &	2	            &	2	            & 2\\
Cores Per socket	&	14	            &	14	            & 16\\
RAM (DDR4)	        &	128 GB	        &	128 GB	        & 128\\
GPU Family          &   Tesla V100      &   Tesla K80       & Quadro RTX 5000\\
GPUs                &   1               &   2               & 4\\
GPU Memory          &   32 GB           &   12 GB           & 16 GB\\
Interconnect	    &	IB-EDR (100G)	&	IB-EDR (100G)	& IB-FDR (56G)\\
\bottomrule
\end{tabular}
\end{table}

\MySubsection{Latency and Throughput Comparison}
\label{sec:results:microbenchmark}

We first perform latency and throughput comparisons between \mpifordask and other communication backends---in particular UCX-Py---using a Ping Pong benchmark. Also we add {\textsf{\small UCX (Tag API)}} and \mvgdr to our comparisons for baseline performance. UCX v1.8.0 was used alongwith {\tt ucx\_perftest}~\cite{ucx-perftest} to evaluate latency and throughput for {\textsf{\small UCX Tag API}}. For \mvgdrns, we used the {\tt osu\_latency} test that is part of the OSU Micro-Benchmark Suite (OMB)~\cite{OMB}. UCX-Py was also evaluated using its own Ping Pong benchmark~\cite{ucxpy-bench-localsr}. UCX-Py has two modes of execution: 1) polling-based, and 2) event-based. The polling-based mode is latency-bound and hence more efficient than the event-based mode. Comparisons also include \mpiforpy running over \mvgdrns---this is labeled as {\textsf{\small MV2-GDR (mpi4py)}}. The benchmark used for \mpiforpy is an in-house Python version of OMB. Lastly we plot \mpifordask that essentially represents the \mpifordask communication device integrated into the Dask Distributed library. We always execute the TCP device using IP over InfiniBand (IPoIB) protocol that provides best performance for this backend. For this reason, we use TCP and IPoIB interchangeably in this section.

\begin{figure*}[htbp]
    \centering
    \vspace{-1.5ex}
    \subfigure[Latency (Small)]
    {
        \label{fig:res:ri2-v100-pingpong-lat-small}
        \includegraphics[width=0.32\textwidth]{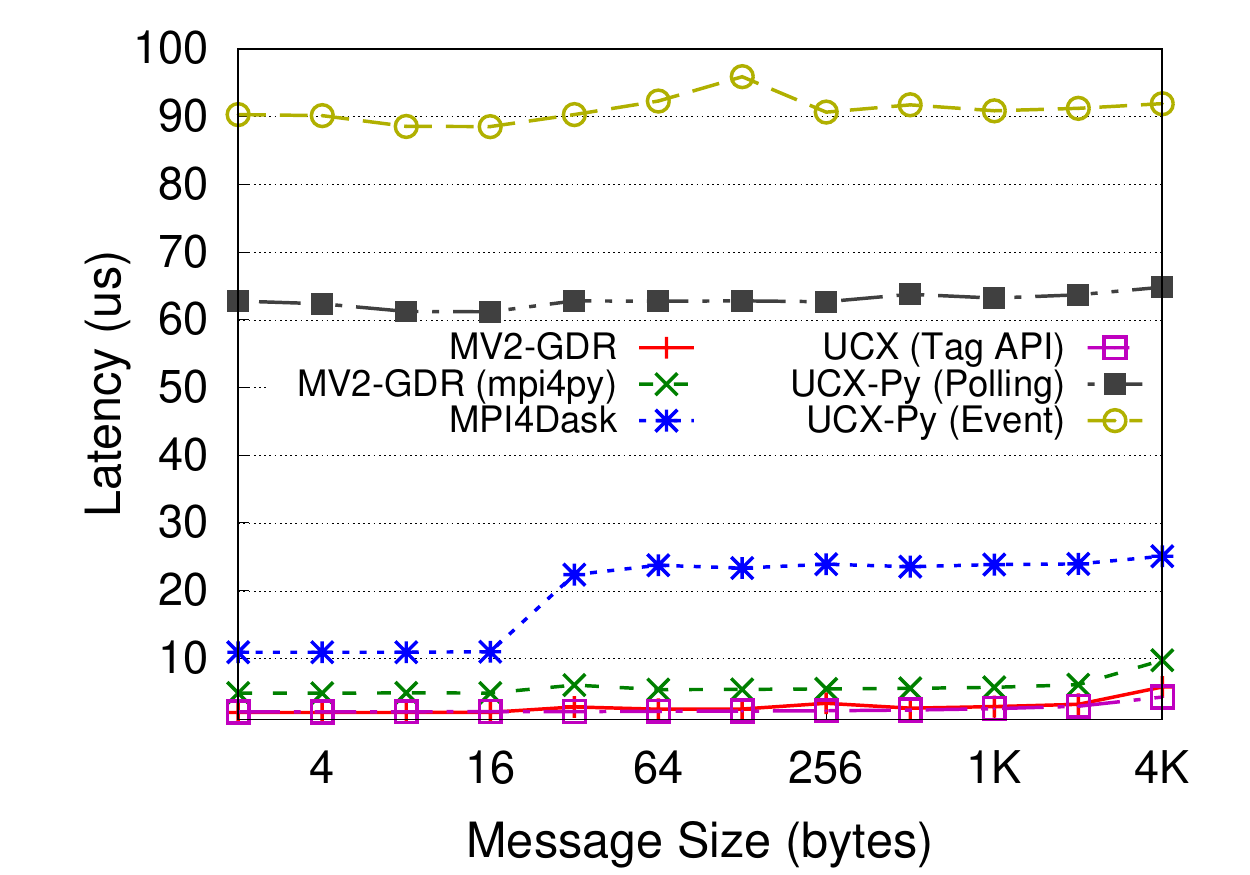}
    }
    \hspace{-3ex}
    \subfigure[Latency (Medium)]
    {
        \label{fig:res:ri2-v100-pingpong-lat-med}
        \includegraphics[width=0.32\textwidth]{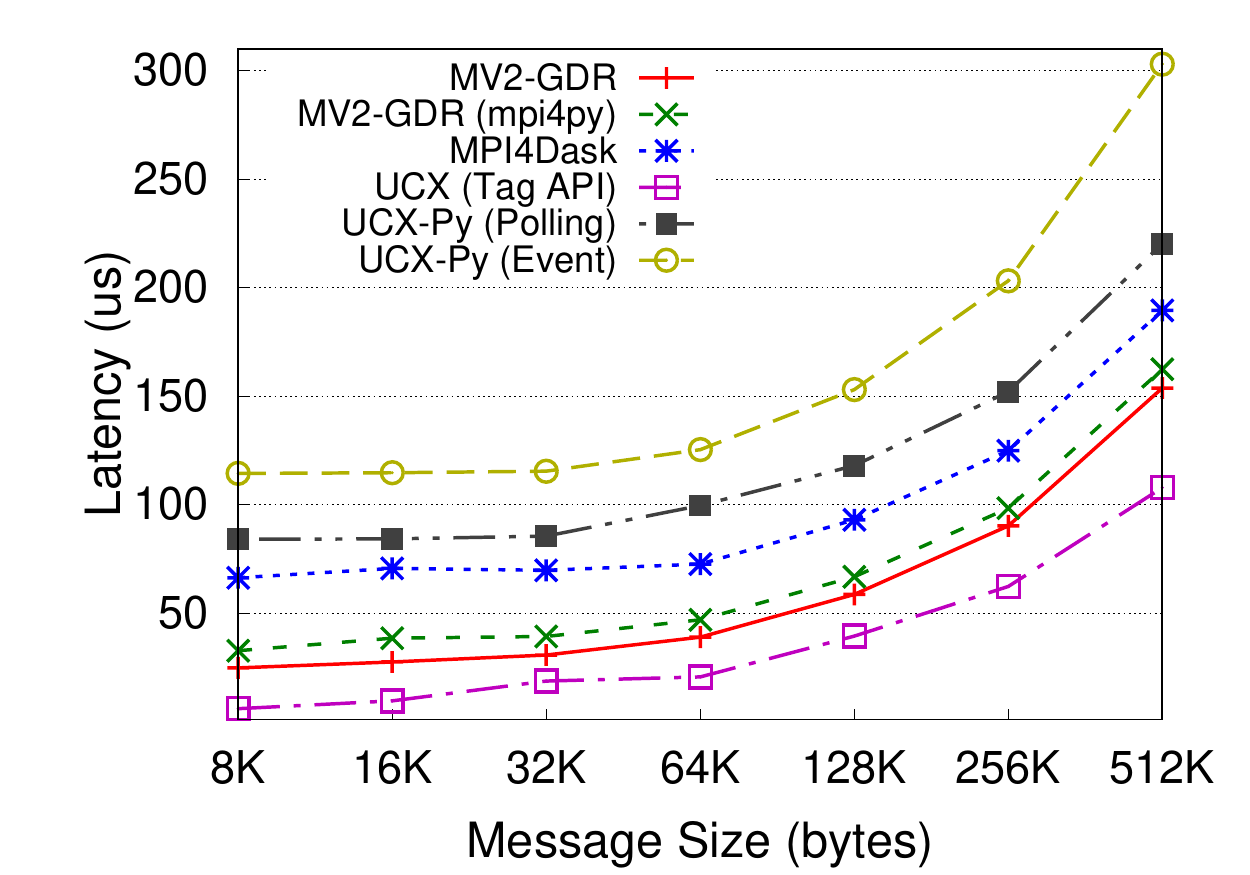}
    }
    \hspace{-3ex}
    \subfigure[Throughput (Large)]
    {
        \label{fig:res:ri2-v100-pingpong-bw-large}
        \includegraphics[width=0.32\textwidth]{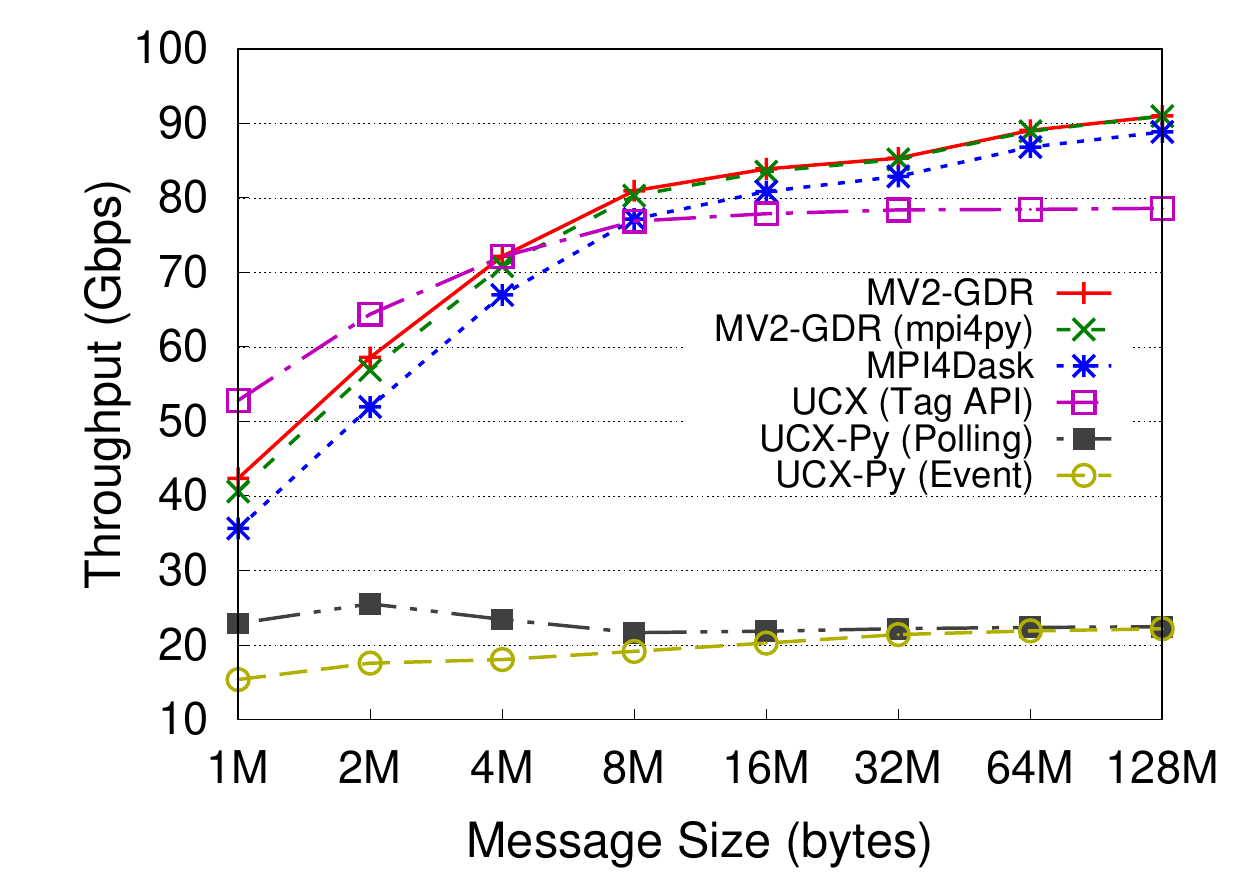}
    }
    \vspace{-1ex}
    \MyCaption{Latency/Bandwidth comparison of \mpifordask with UCX-Py (Polling and Event Modes) using Ping Pong Benchmark on RI2 Cluster with V100 GPUs. {\textsf{\small UCX (Tag API)}} and \mvgdr numbers are also presented for baseline performance.}
    \vspace{-2ex}
    \label{fig:res:ri2-v100-pingpong-latbw}
\end{figure*}

 \vspace{1ex}

\begin{figure*}[htbp]
    \centering
    \vspace{-1.5ex}
    \subfigure[Latency (Small)]
    {
        \label{fig:res:ri2-k80-pingpong-lat-small}
        \includegraphics[width=0.32\textwidth]{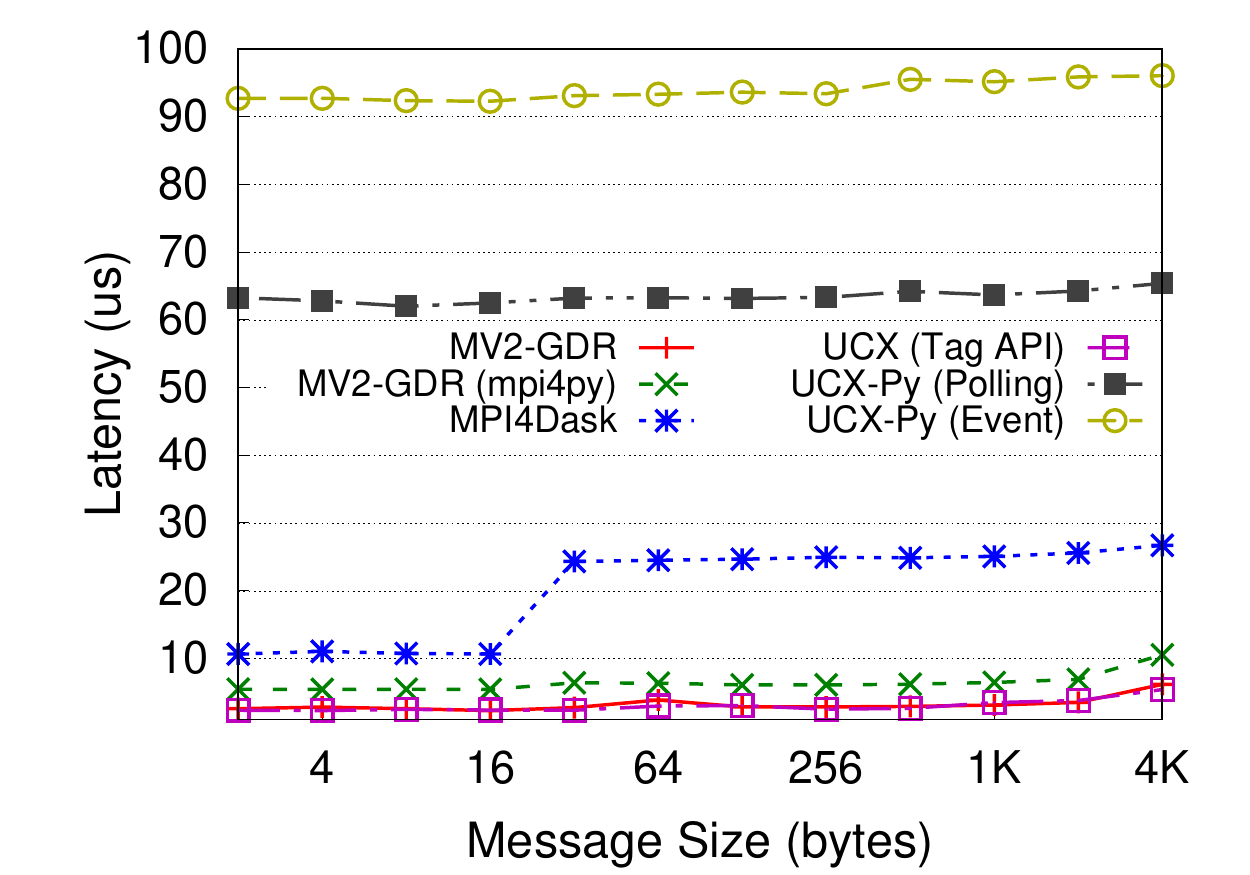}
    }
    \hspace{-3ex}
    \subfigure[Latency (Medium)]
    {
        \label{fig:res:ri2-k80-pingpong-lat-med}
        \includegraphics[width=0.32\textwidth]{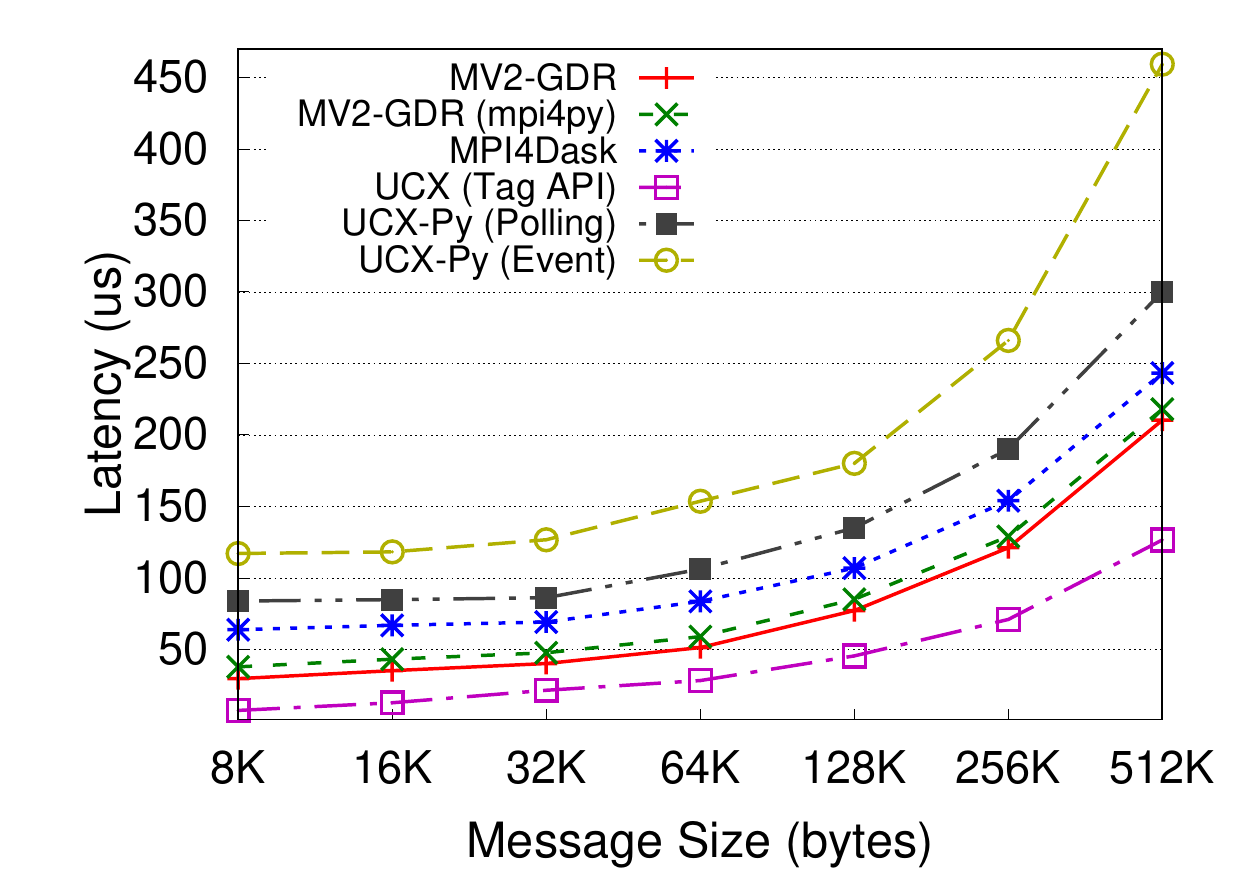}
    }
    \hspace{-3ex}
    \subfigure[Throughput (Large)]
    {
        \label{fig:res:ri2-k80-pingpong-bw-large}
        \includegraphics[width=0.32\textwidth]{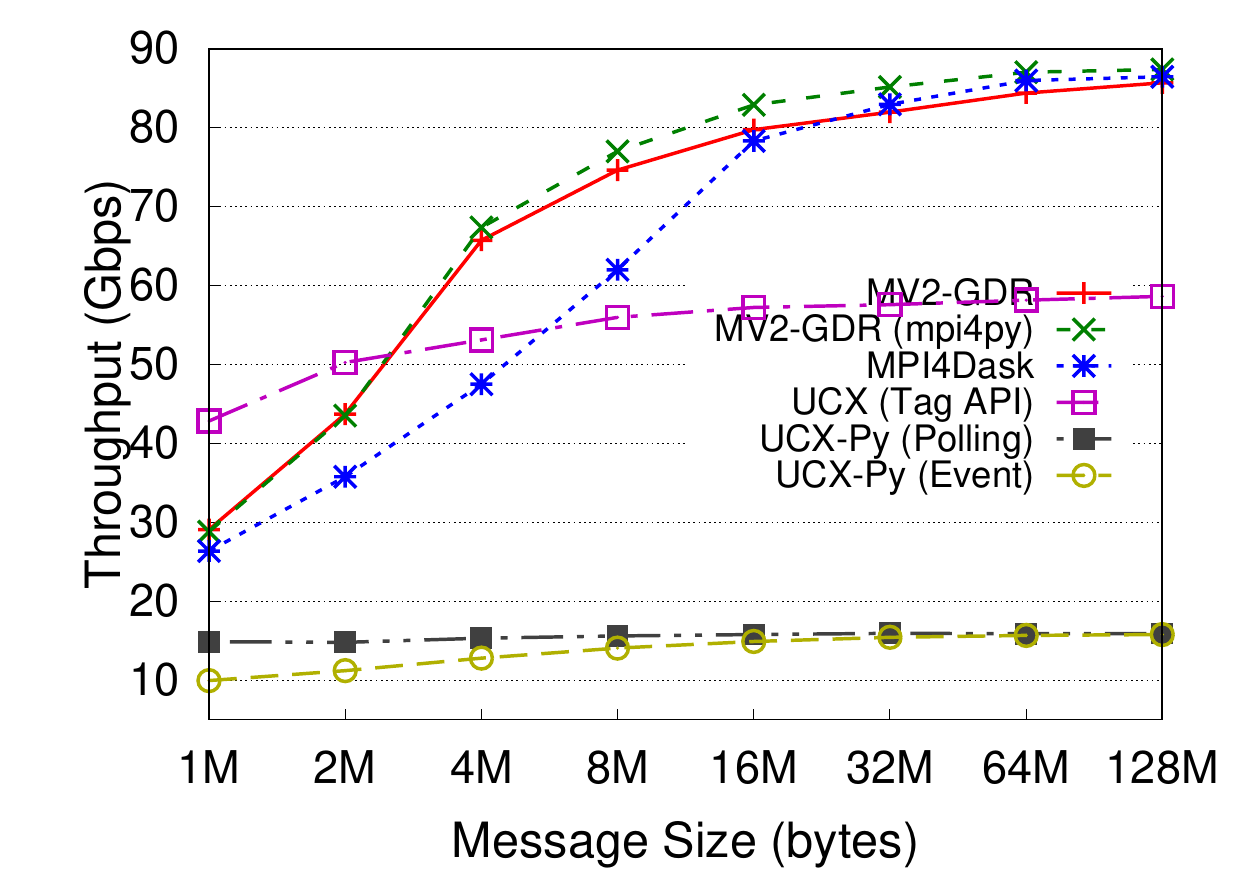}
    }
    \vspace{-1ex}
    \MyCaption{Latency/Throughput comparison of \mpifordask with UCX-Py (Polling and Event Modes) using Ping Pong Benchmark on RI2 Cluster with K80 GPUs. {\textsf{\small UCX (Tag API)}} and \mvgdr numbers are also presented for baseline performance.}
    \vspace{-2ex}
    \label{fig:res:ri2-k80-pingpong-latbw}
\end{figure*}

Figure~\ref{fig:res:ri2-v100-pingpong-lat-small} shows latency for message sizes $1$ Byte to $4$ KByte---here \mpifordask outperforms {\textsf{\small UCX-Py (Polling)}} by $5\times$. Similarly for medium-sized messages presented in Figure~\ref{fig:res:ri2-v100-pingpong-lat-med}---$8$ KByte to $512$ KByte---\mpifordask is better than {\textsf{\small UCX-Py (Polling)}} by $2-3\times$. In throughput comparisons for large messages, $1$ MByte and beyond, \mpifordask outperforms {\textsf{\small UCX-Py (Polling)}} by $3-4\times$ in throughput as shown in Figure~\ref{fig:res:ri2-v100-pingpong-bw-large}. It is important to note that the difference in performance between {\textsf{\small MV2-GDR (mpi4py)}} and {\textsf{\small MV2-GDR}} is minimal. This means that regular Python functions over native code do not introduce much overhead. However when executing communication coroutines---by employing \mpifordask or UCX-Py---there is additional overhead due to the {\tt asyncio} framework. We observe similar performance patterns for Tesla K80 GPUs in latency and throughput comparisons between \mpifordask and {\textsf{\small UCX-Py (Polling)}} as depicted in Figure~\ref{fig:res:ri2-k80-pingpong-latbw}. As noted earlier, we are focusing on performance comparison of \mpifordask with {\textsf{\small UCX-Py (Polling)}} since it is the more efficient mode of the UCX-Py library. 

\MySubsection{Application Benchmarks}
\label{sec:results:appbenchmark}

We also evaluated \mpifordask against UCX-Py for two application benchmarks: 1) sum of cuPy array and its transpose, and 2) cuDF merge. The cuPy benchmark presents strong scaling results as the problem size remains the same as more Dask workers are used in the computation. On the other hand, the cuDF benchmark presents weak scaling results as the problem size increases with increase in the number of Dask workers. This evaluation was done on the in-house RI2 cluster.

\MySubsubsection{Sum of cuPy Array and its Transpose}

This benchmark~\cite{dask-app-transpose-sum} creates a cuPy array and distributes its chunks across Dask workers. The benchmark adds these distributed chunks to their transpose, forcing the GPU data to move around over the network. The following operations are performed:\\
\indentcentercol \texttt{y = x + x.T}\\
\indentcentercol \texttt{y = y.persist()}\\
\indentcentercol \texttt{wait(y)}

\begin{figure*}[htbp]
    \centering
    \subfigure[Execution Time Comparison]
    {
        \label{fig:res:ri2-app1-cupy-exectime}
        \includegraphics[width=0.3\textwidth]{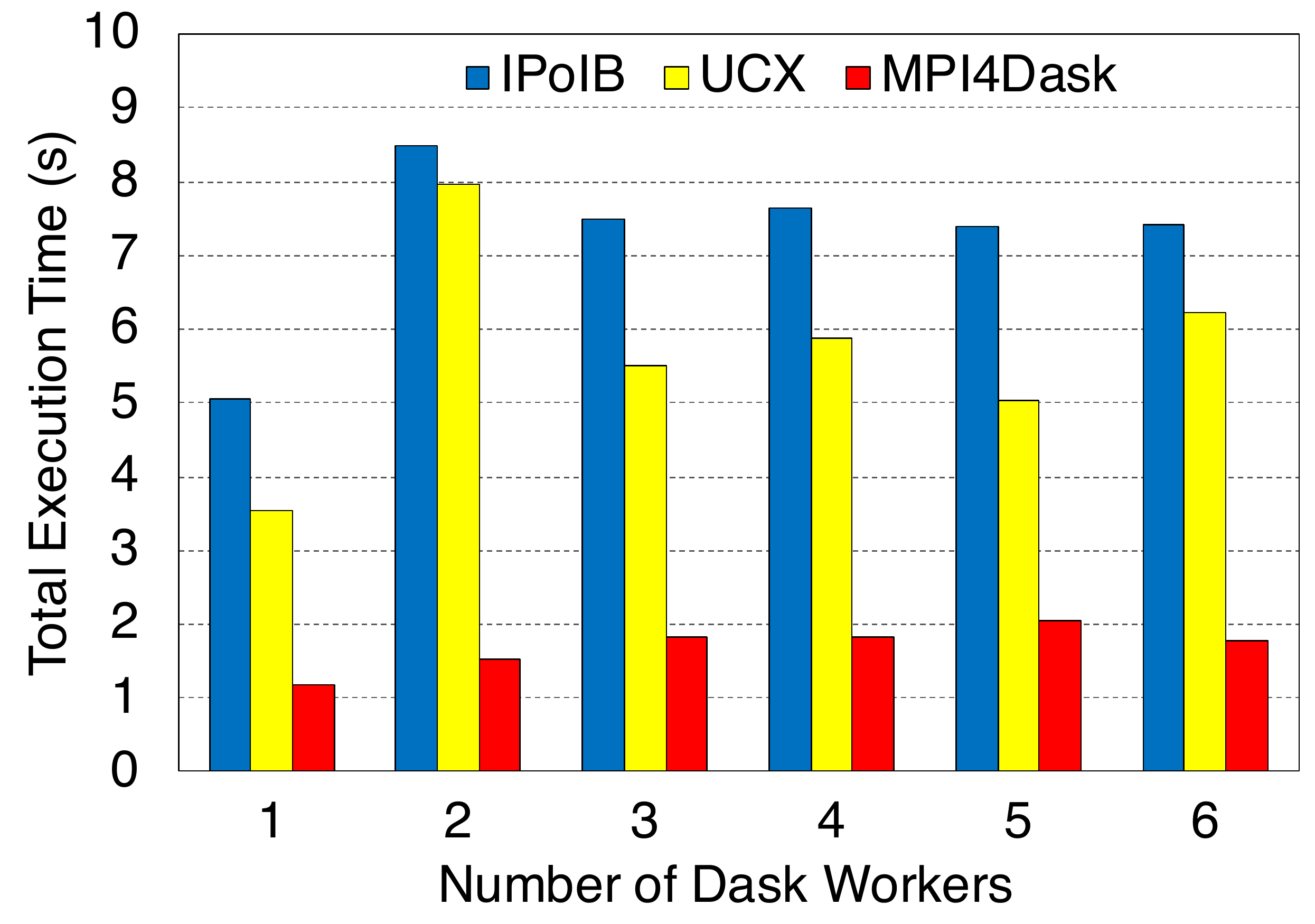}
    }
    \hspace{-0ex}
    \subfigure[Communication Time Comparison]
    {
        \label{fig:res:ri2-app1-cupy-commtime}
        \includegraphics[width=0.3\textwidth]{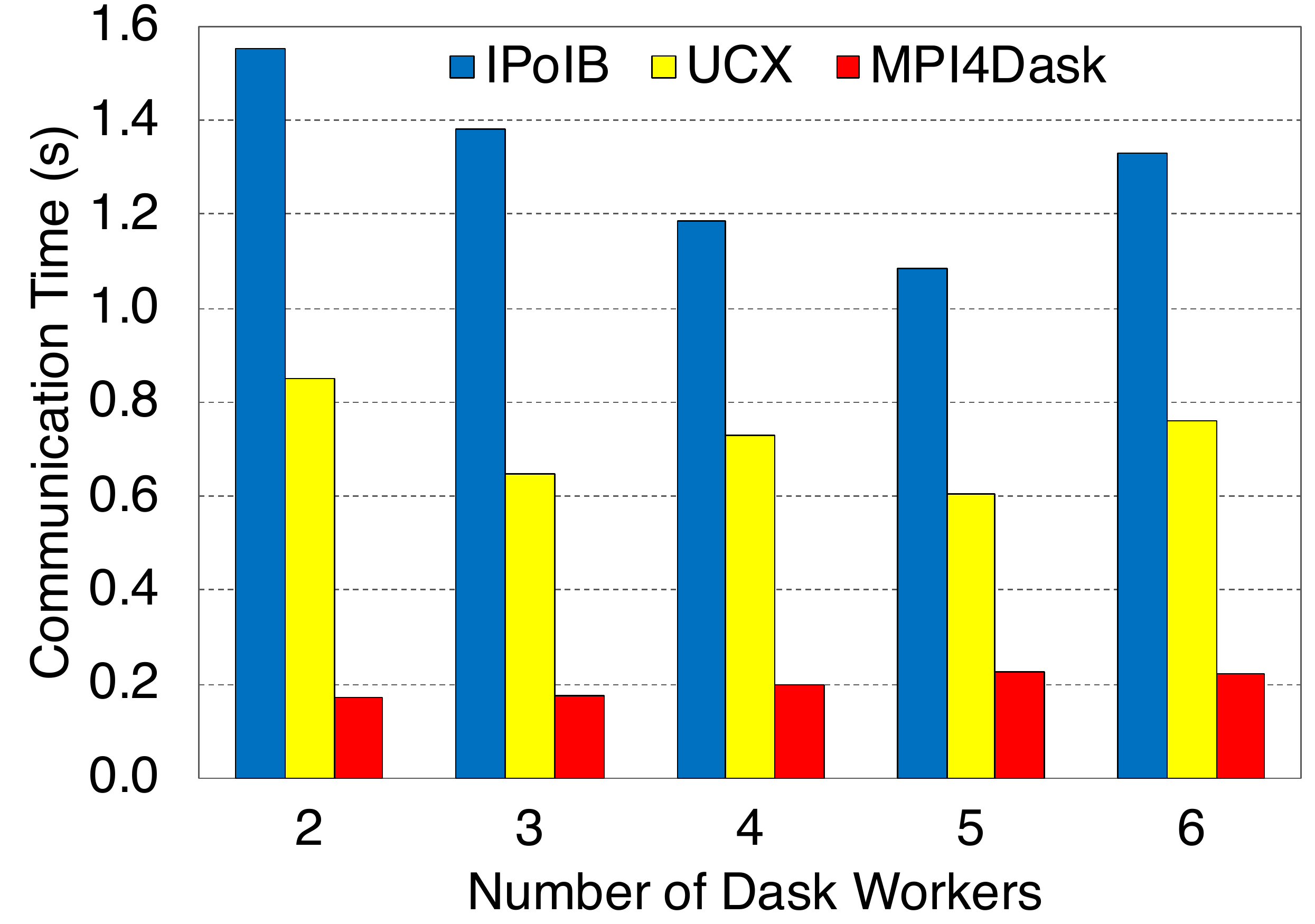}
    }
    \hspace{-0ex}
    \subfigure[Aggregate Throughput Between Workers]
    {
        \label{fig:res:ri2-app1-cupy-bw}
        \includegraphics[width=0.3\textwidth]{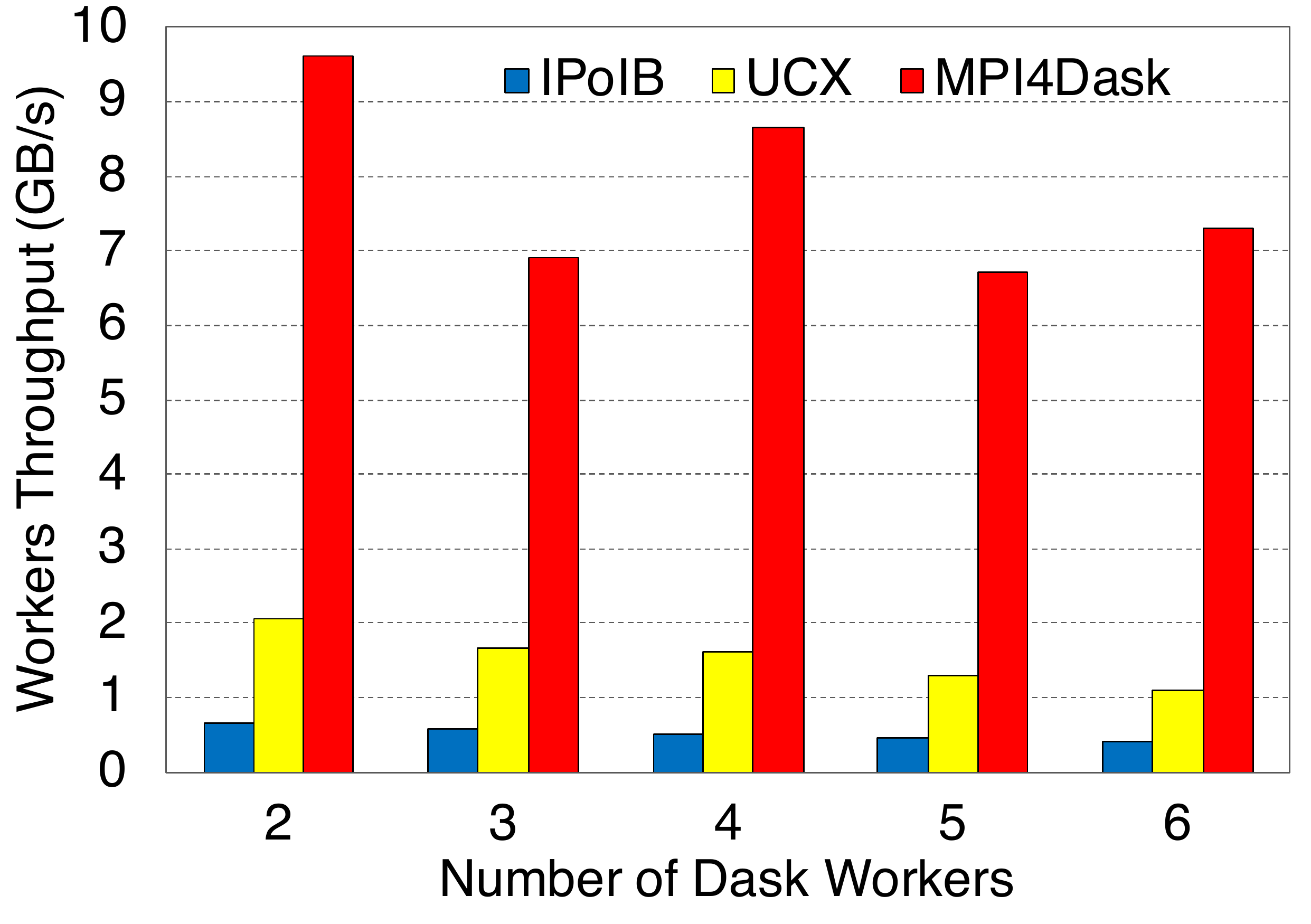}
    }
    \vspace{-1ex}
    \MyCaption{Sum of cuPy Array and its Transpose (cuPy Dims: 16K$\times$16K, Chunk size: 4K, Partitions: 16): Performance Comparison between IPoIB, UCX, and \mpifordask on the RI2 Cluster. This benchmark presents strong scaling results. $28$ threads are started in a single Dask worker.}
    \vspace{-1ex}
    \label{fig:res:ri2-app1-cupy}
\end{figure*}

Performance comparison graphs for the first application benchmark---sum of cuPy array with its transpose---are shown in Figure~\ref{fig:res:ri2-app1-cupy}. Dask follows the One Process per GPU (OPG) model of execution, which means that a single worker process with multiple threads is initiated for each GPU. On the RI2 cluster, each node has a single GPU. For this reason, we instantiate a single worker process with $28$ worker threads to fully exploit the available CPU cores. Figure~\ref{fig:res:ri2-app1-cupy-exectime} shows execution time where we are witnessing an average speedup of $3.47\times$ for $2-6$ Dask workers. Figure~\ref{fig:res:ri2-app1-cupy-commtime} shows the communication time for the benchmark run. This shows that \mpifordask is better than UCX by $6.92\times$ on average for $2-6$ workers. Average throughput comparison for Dask workers is shown in Figure~\ref{fig:res:ri2-app1-cupy-bw}, which depicts that \mpifordask outperform UCX by $5.17\times$ on average for $2-6$ workers. This benchmark application is presenting strong scaling results. As can be seen in Figure~\ref{fig:res:ri2-app1-cupy-exectime}, the cuPy application does not exhibit impressive speedups as the number of workers are increased. This is due to the nature of this benchmark that is designed to stress and evaluate communication performance. This is suitable for this paper because we are primarily interested in comparative performance of communication devices and not demonstrating application-level performance of the Dask ecosystem.

\begin{figure*}[htbp]
    \centering
    \subfigure[Execution Time Comparison]
    {
        \label{fig:res:ri2-app2-cudf-exectime}
        \includegraphics[width=0.3\textwidth]{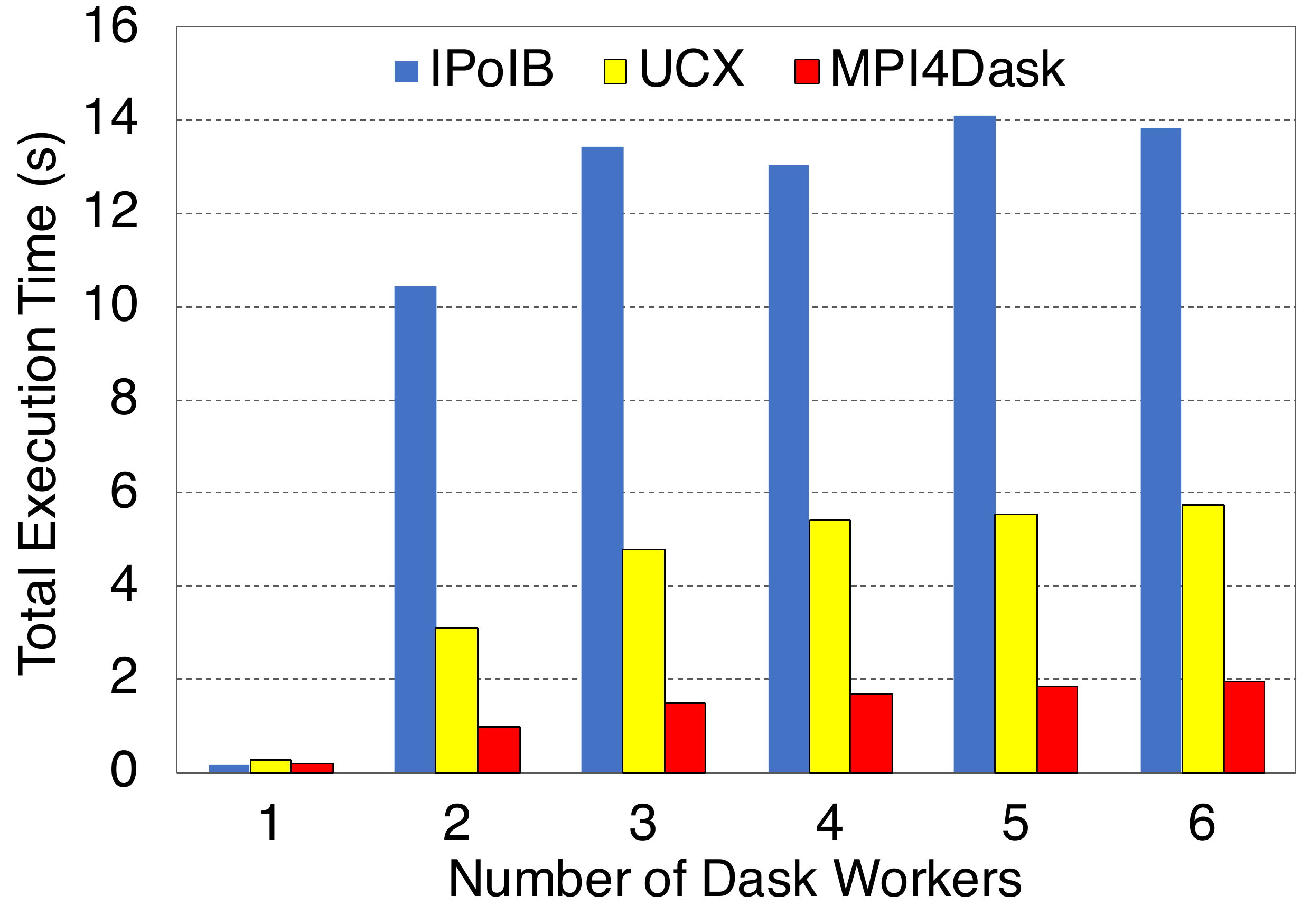}
    }
    \hspace{-0ex}
    \subfigure[Communication Time Comparison]
    {
        \label{fig:res:ri2-app2-cudf-commtime}
        \includegraphics[width=0.3\textwidth]{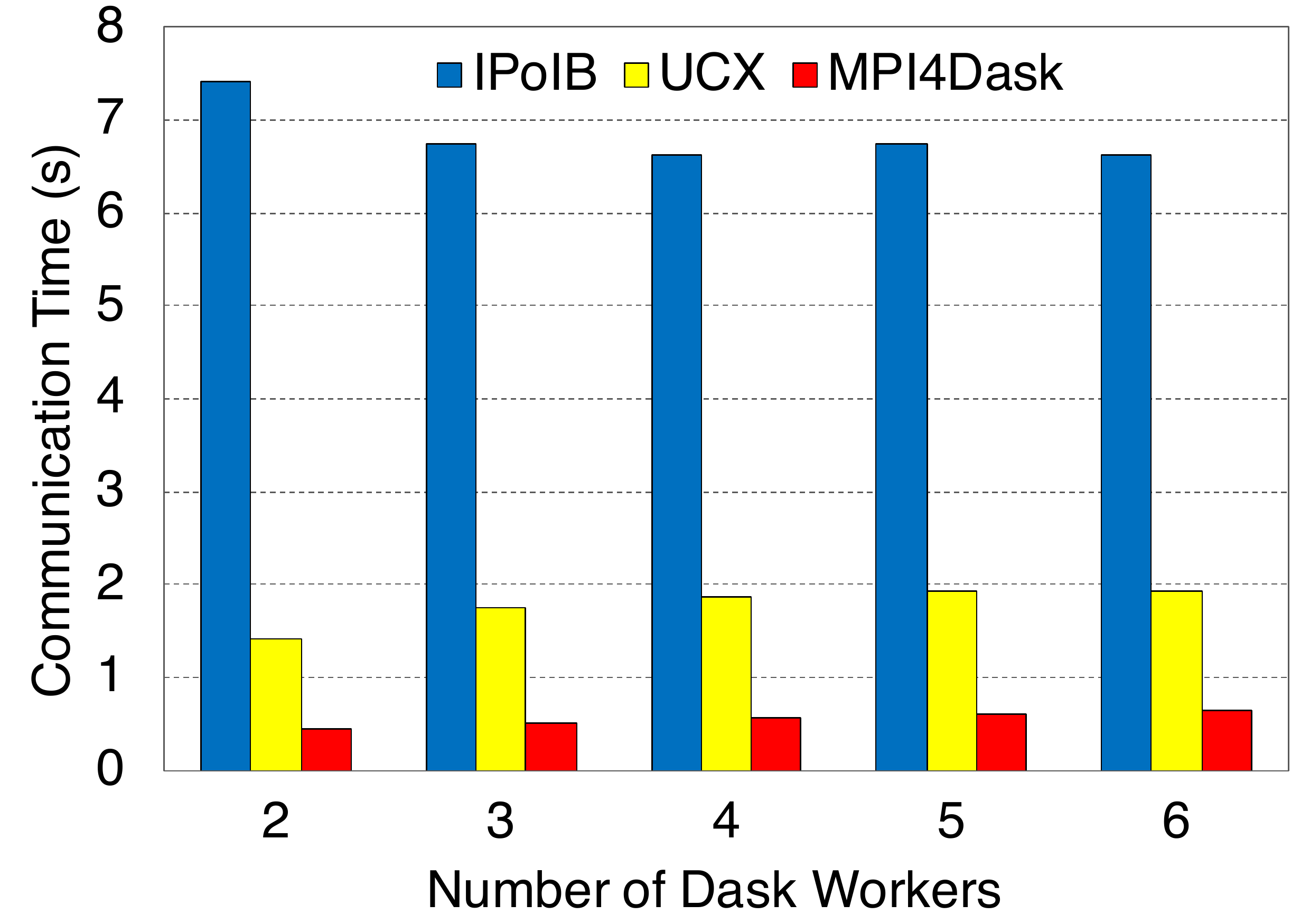}
    }
    \hspace{-0ex}
    \subfigure[Aggregate Throughput Between Workers]
    {
        \label{fig:res:ri2-app2-cudf-bw}
        \includegraphics[width=0.3\textwidth]{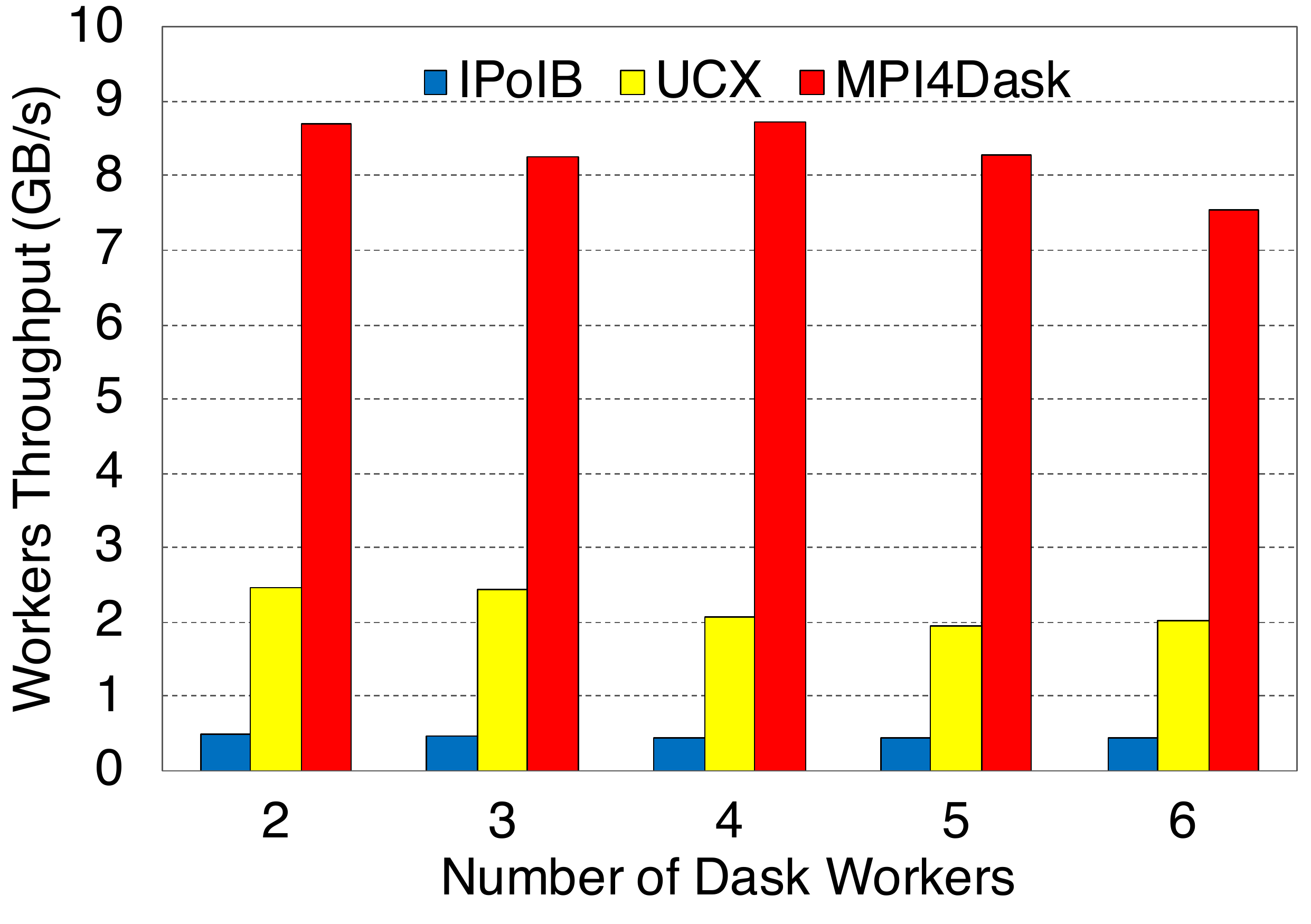}
    }
    \vspace{-1ex}
    \MyCaption{cuDF Merge Operation Benchmark (Chunk Size: 1E8, Shuffle: True, Fraction Match: 0.3): Performance Comparison between IPoIB, UCX, and \mpifordask on the RI2 Cluster. This benchmark presents weak scaling results. $28$ threads are started in a single Dask worker. $3.2$ GB of data is processed per Dask worker (or GPU)}
    \vspace{-1ex}
    \label{fig:res:ri2-app2-cudf}
\end{figure*}

\MySubsubsection{cuDF Merge}

cuDF {\tt DataFrame}s are table-like data-structure that are stored in the GPU memory. As part of this application~\cite{dask-app-cudf-merge}, a merge operation is carried out for multiple cuDF data frames. Performance comparison graphs for the second application benchmark---cuDF merge operation---are shown in Figure~\ref{fig:res:ri2-app2-cudf}. Figure~\ref{fig:res:ri2-app2-cudf-exectime} shows execution time where we are witnessing an average speedup of $3.11\times$ for $2-6$ Dask workers. Figure~\ref{fig:res:ri2-app2-cudf-commtime} shows time for communication that is taking place. This shows that \mpifordask is better than UCX by $3.22\times$ on average for $2-6$ workers. Average throughput comparison for Dask workers is presented in Figure~\ref{fig:res:ri2-app2-cudf-bw}, which depicts that \mpifordask outperforms UCX by $3.82\times$ on average for $2-6$ workers. The cuDF merge benchmark is presenting weak scaling results. As can be observed in Figure~\ref{fig:res:ri2-app2-cudf}, the execution time with \mpifordask is increasing slightly with an increase in the number of workers. This is due to efficient communication performance provided by \mpifordask to the Dask execution. Note that $3.2$, $6.4$, $9.6$, $12.8$, $16$, and $19.2$ GB of data is processed for $1$, $2$, $3$, $4$, $5$, and $6$ Dask workers respectively.

Note that UCX-Py is executed in polling-mode that is the efficient mode. Reasons for better overall performance of \mpifordask against other counterparts include better point-to-point performance of \mvgdr, chunking implemented for messages greater than $1$ GB, and efficient coroutine implementation for \mpifordask as compared to UCX-Py. Unlike UCX-Py that implements a separate coroutine to make communication progress for UCX worker, \mpifordask ensures {\em cooperative} progression where every communication coroutine triggers the communication progression engine.

\MySubsection{Scalability Results on the TACC Frontera (GPU) Cluster}
\label{sec:results:scalability}
\vspace*{-0.5ex}

\begin{figure*}[htbp]
    \centering
    \subfigure[Execution Time Comparison for Sum of cuPy Array and its Transpose Benchmark. (cuPy Dims: 20E3$\times$20E3, Chunk size: 5E2, Partitions: 16E2)]
    {
        \label{fig:res:frontera-cupy-exectime}
        \includegraphics[width=0.3\textwidth]{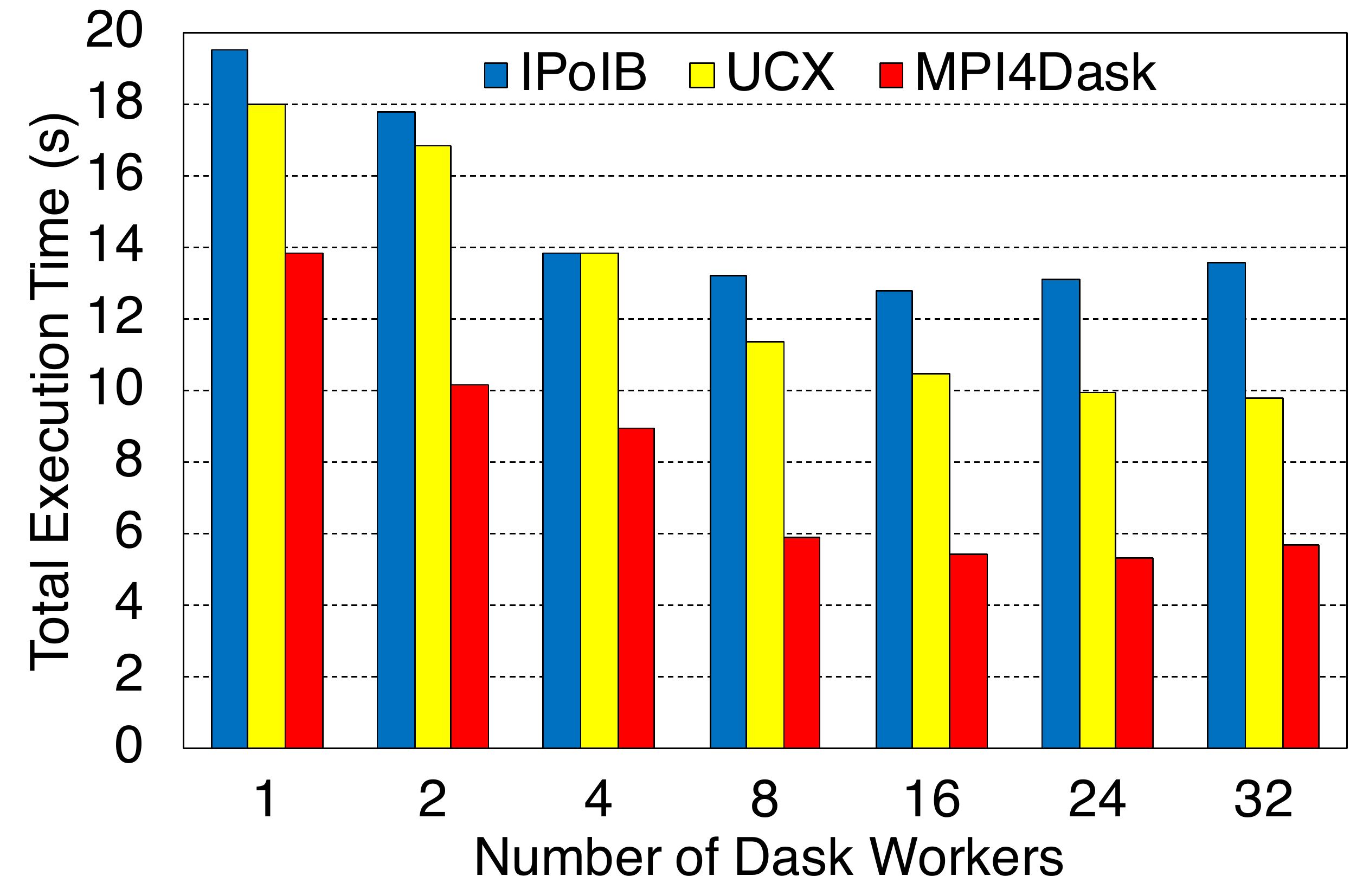}
    }
    \hspace{-0ex}
    \subfigure[Execution Time Comparison for cuDF Merge Benchmark. (Chunk Size: 5E7, Shuffle: True, Fraction Match: 0.3)]
    {
        \label{fig:res:frontera-cudf-exectime}
        \includegraphics[width=0.3\textwidth]{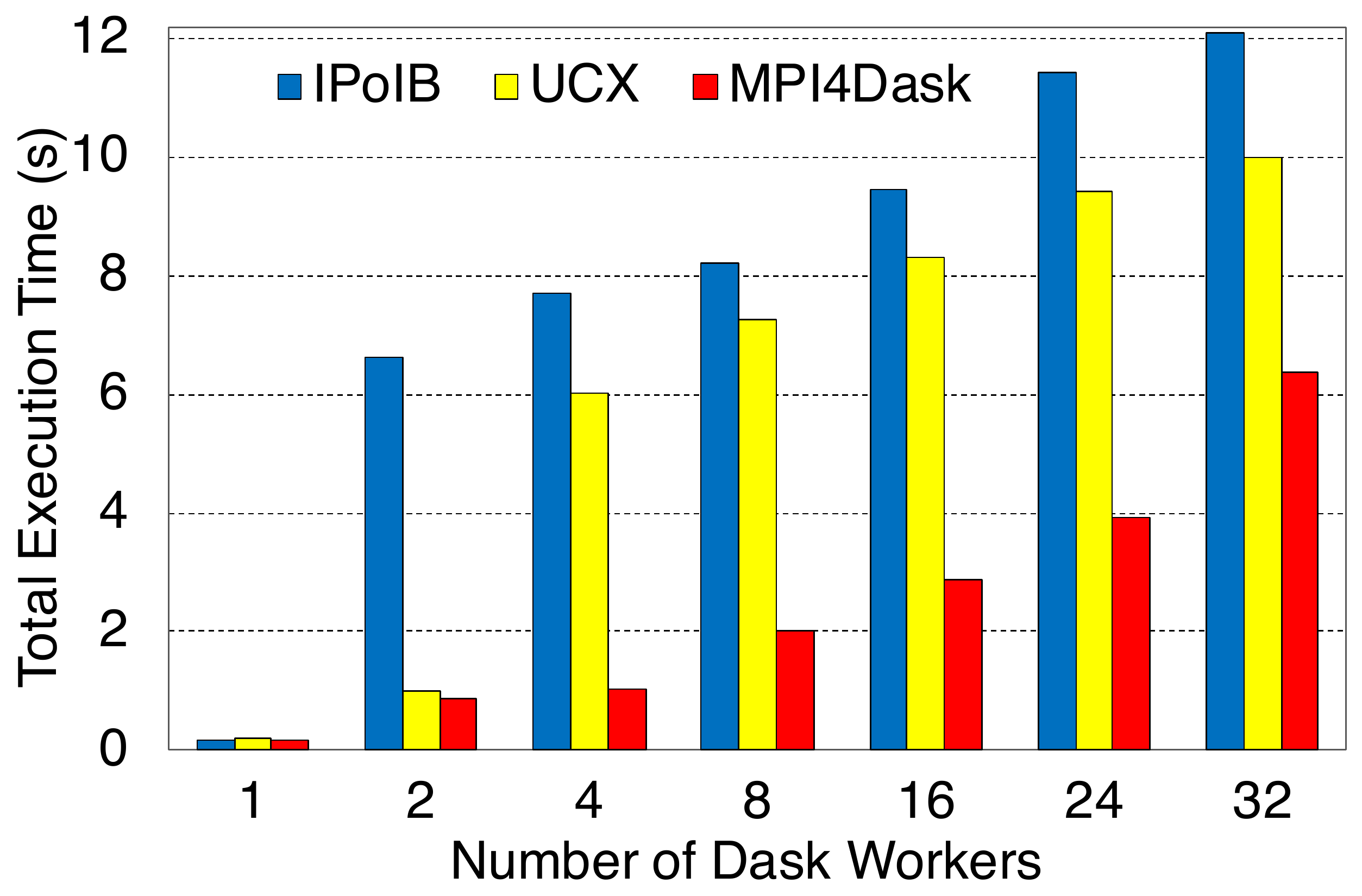}
    }
    \hspace{-0ex}
    \subfigure[Average Throughput For Workers for the cuDF Merge Benchmark]
    {
        \label{fig:res:frontera-cudf-throughput}
        \includegraphics[width=0.3\textwidth]{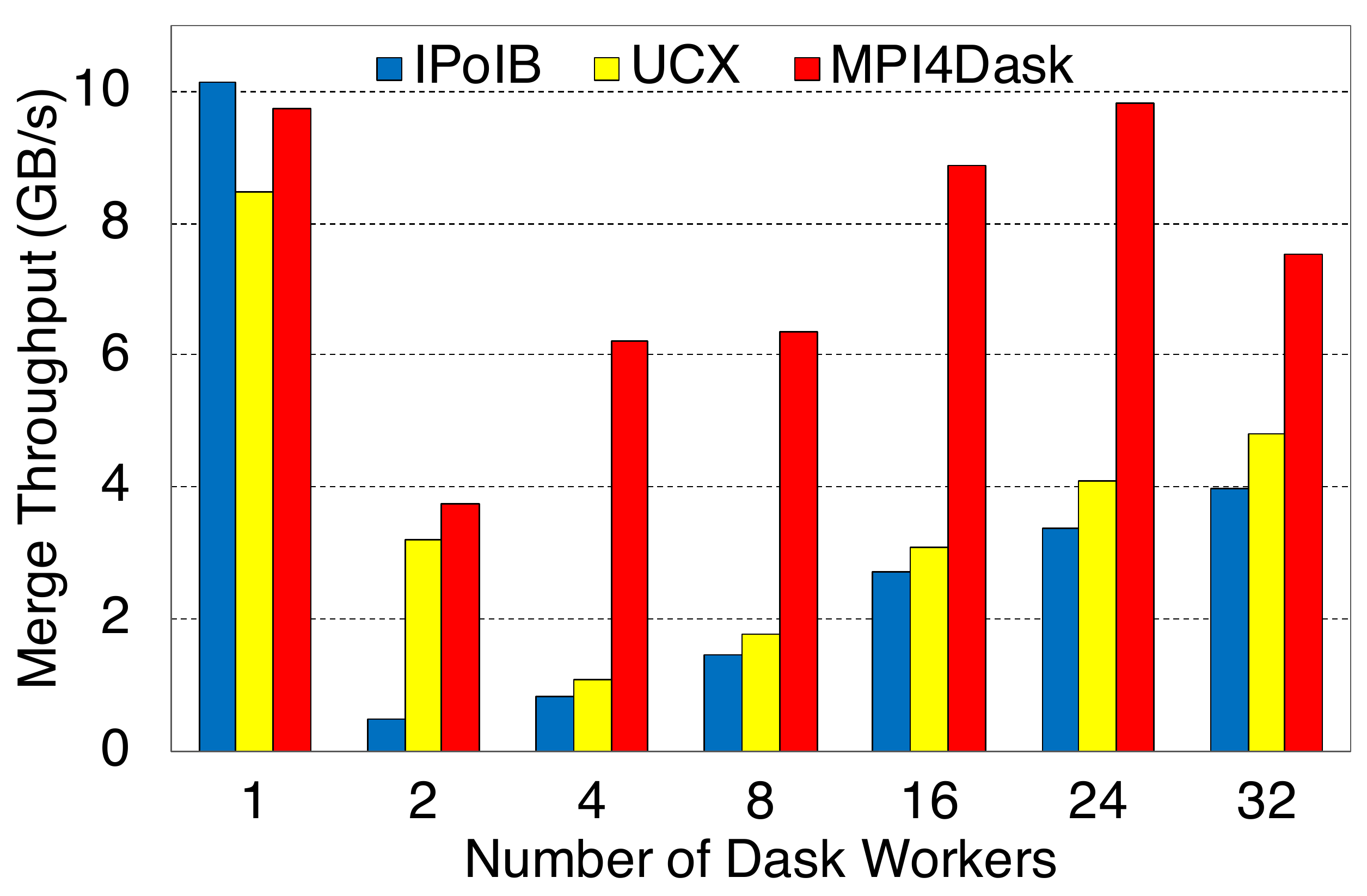}
    }
    \vspace{-1ex}
    \MyCaption{Execution Time Comparison for Sum of cuPy Array with its Transpose Benchmark, cuDF Merge Benchmark, and Average Throughput for Dask Workers for the cuDF Merge Benchmark. This evaluation was done on the Frontera (GPU) system at TACC. This comparison is between \mpifordask, UCX, and TCP (using IPoIB) communication devices. There are $4$ GPUs in a single node. Each Dask worker has $8$ threads.}
    \vspace{-1ex}
    \label{fig:res:frontera-app}
\end{figure*}

This section presents the scalability results for the two application benchmarks introduced earlier in Section~\ref{sec:results:appbenchmark}. This evaluation was done on the Frontera (GPU) system that is equipped with $360$ NVIDIA Quadro RTX 5000 GPUs in $90$ nodes. Following the OPG model of execution, we started $4$ processes on a single node---one process per GPU---with $8$ worker threads in each process. This evaluation is presented in Figure~\ref{fig:res:frontera-app}. Figure~\ref{fig:res:frontera-cupy-exectime} shows the execution time comparison between TCP (using IPoIB), UCX, and \mpifordask devices for the sum of cuPy with its transpose application with $1-32$ Dask workers---here \mpifordask outperforms UCX by an average factor of $1.71\times$. Figures~\ref{fig:res:frontera-cudf-exectime} and~\ref{fig:res:frontera-cudf-throughput} shows the execution time and throughput comparison between all communication devices for the cuDF merge operation. Again, \mpifordask outperforms UCX by an average factor of $2.91\times$ for overall execution time for $1-32$ Dask workers. The merge operation throughout depicts the rate at which workers merge the cuDF input data and it also follows a similar pattern. The performance results shown in Figure~\ref{fig:res:frontera-cupy-exectime} present strong scaling results for the cuPy application, while Figures~\ref{fig:res:frontera-cudf-exectime} and~\ref{fig:res:frontera-cudf-throughput} present weak scaling results for the cuDF merge operation. The performance shown by $1$ Dask worker in Figures~\ref{fig:res:frontera-cudf-exectime} and~\ref{fig:res:frontera-cudf-throughput} is efficient due to small problem size and no communication overhead. However, the problem size becomes larger and more realistic with additional Dask workers. Note that $1.6$ GB of data is processed per GPU for the cuDF application---this means that $51.2$ GB of data is processed on $32$ GPUs.

\MySection{Related Work}
\label{sec:background:related}

A framework similar to Dask is the Apache Spark software~\cite{ZahariaXinEtAl16cacm}. It supports programming data science applications using Scala, Java, Python and R. Traditionally Spark has also only supported execution on hosts (CPUs). This meant that it was not able to exploit massive parallelism offered by GPUs. However GPU support has been added by the RAPIDS project recently to the Apache Spark 3.0 through a Spark-RAPIDS plugin~\cite{Spark-RAPIDS}. Vanilla versions of Apache Spark only support data communication over Ethernet or other high-speed networks using the TCP backend. There have been efforts~\cite{RDMASparkOSU, SparkRDMAMellanox} to address this shortcoming for Apache Spark. 

The Dask Distributed library currently support two communication backends. The first device---called TCP---makes use of Python's Tornado framework~\cite{dask-tornado}. The second device---called UCX---uses a Cython wrapper called UCX-Py~\cite{ucx-py} over the UCX~\cite{UCX} communication library. In an earlier effort~\cite{ShafiHiPC20}, we have developed a new communication device called Blink for Dask. However Blink is limited to CPU-only execution of Dask programs and hence is not relevant here. Also, recently there has been another effort~\cite{MPICommGitHubIssue} to develop an MPI communication layer on the Dask Distributed github repository. However, this is currently a work in progress and is not addressing GPU-based data science applications, which is the focus of this paper. At this time, it is only possible to use TCP and UCX based devices for Dask on cluster of GPUs. This paper demonstrates that \mpifordask---developed as part of this work---delivers better performance than the current state of the art communication options for Dask on cluster of GPUs.
\MySection{Conclusions and Future Work}
\label{sec:conclusion}
\vspace*{-0.5ex}

Python is an emerging language on the landscape of scientific computing with support for distributed computing engines like Dask. Rivaling the Apache Spark ecosystem, Dask allows incremental parallelization of Big Data applications. Recently the Dask software has been extended, as part of the NVIDIA RAPIDS framework, to support distributed computation on cluster of GPUs. Support for GPU-based data storage and processing APIs like cuPy and cuDF---GPU-counterparts of numPy and Pandas respectively---has also been added. As part of this work, we have extended the Dask Distributed library with a new communication device called \mpifordask based on the popular MPI standard. \mpifordask exploits the \mpiforpy wrapper software on top of the \mvgdr library to offer an efficient alternative to existing backends---TCP and UCX---from within a non-blocking asynchronous I/O framework. This is done by implementing high-performance communication coroutines using MPI that support the {\tt async}/{\tt await} syntax. The performance evaluation done, on an in-house cluster built with Tesla V100/K80 GPUs and the Frontera (GPU) cluster equipped with Quadro RTX 5000 GPUs---using the Ping Pong micro-benchmark and two other application benchmarks---suggest that \mpifordask clearly outperforms other communication devices. This is due to efficient implementation of communication coroutines in \mpifordask that use the {\em cooperative} progress approach with non-blocking point-to-point MPI calls. In the future we plan to extend \mpifordask with support for dynamic process management, which will allow entities like workers and clients to dynamically join and leave the Dask Cluster. Also we aim to extend \mpifordask for host-based (CPU only) distributed computation with Dask.

\vspace*{-0.3ex}

\balance

\bibliographystyle{IEEEtran}
\bibliography{HiPC20/references/blink.bib}

\end{document}